\documentclass[11pt]{article}

\usepackage{graphicx,epstopdf,float,multirow,booktabs}
\usepackage{amsmath,amssymb,amsfonts,amsthm}
\usepackage{bbm,bm,bbold}
\usepackage{url}

\usepackage{xcolor}
\usepackage{enumitem}
\usepackage{tikz}
\usetikzlibrary{calc}
\usetikzlibrary{patterns,snakes}
\usepackage{setspace,color}
\usepackage{lineno}
\usepackage{mathtools}
\usepackage{caption}
\usepackage{subcaption}
\usepackage{comment}

\usepackage{geometry}
\usepackage{hyperref}
\hypersetup{
    colorlinks=true,
    linkcolor=blue,
    filecolor=magenta,
    urlcolor=blue,
    citecolor=blue,
    pdfauthor={},       
    pdftitle={Mortality Heterogeneity and Actuarial Fairness in China's NDC Pension System},
    pdfsubject={},
}
\usepackage[sort&compress]{natbib}
\usepackage{cleveref}

\DeclareMathOperator*{\argmin}{arg\,min}

\begin{document}
\doublespace

\title{\Large\textbf{Mortality Heterogeneity and Actuarial Fairness in China's Notional Defined Contribution Pension System}}

\author{
   Xiaoyu Dong\thanks{University of Illinois Urbana-Champaign, Champaign, USA. Email: xd23@illinois.edu} \and
   Hong Li\thanks{University of Guelph, Guelph, Canada. Email: lihong@uoguelph.ca} \and
   Kenneth Q. Zhou\thanks{University of Waterloo, Waterloo, Canada. Email: kenneth.zhou@uwaterloo.ca} \and
   Xiaobai Zhu\thanks{The Chinese University of Hong Kong, Hong Kong, China. Email: xiaobaizhu58@cuhk.edu.hk}
}


\maketitle
\thispagestyle{empty}

\begin{abstract}
We study actuarial fairness in China’s notional defined contribution (NDC) pension system when mortality differs across income groups. Under current rules, individual account balances are converted into monthly benefits using an official annuity divisor that depends only on retirement age. We develop a mortality-differentiated Lee--Carter framework with group-specific baseline mortality schedules and a common period effect, estimated by combining national mortality data for 1994--2020 with CHARLS subgroup data for 2011--2020. To model cross-group mortality parsimoniously under limited data, we parameterize the baseline schedules using Hermite splines. Applying the model to China’s NDC system, we find substantial actuarial unfairness in the current age-only divisor. The subsidy rises monotonically with income, implying both an aggregate actuarial shortfall and a reverse transfer from poorer to richer retirees. We then compare four implementable income-dependent annuitization rules, ranging from a simple bracket design to marginal-rule alternatives, and show that all substantially reduce the reverse transfer.
\end{abstract}
	
\noindent{{\bf Keywords and phrases}: Actuarial fairness; Hermite spline; Mortality heterogeneity; Notional defined contribution; Lee--carter model.

\newpage

\section{Introduction}

China's urban employee pension scheme converts the balance in an individual's notional account into a stream of monthly benefits by dividing the account balance at retirement by an official annuity divisor, commonly called the ``counting month'' in Chinese policy practice. Under the current rule, this divisor depends only on retirement age. When mortality differs systematically by socioeconomic status, such an age-only divisor cannot be actuarially neutral: longer-lived individuals receive a larger expected present value of benefits, and the combination of higher longevity and larger individual accounts can generate an implicit redistribution from poorer to richer retirees. This paper studies that problem for China's notional defined contribution (NDC) pension system. We develop a mortality-differentiated Lee--Carter framework with group-specific mortality predictions, estimate it by combining national mortality data with the China Health and Retirement Longitudinal Study (CHARLS)\footnote{\url{https://charls.pku.edu.cn/en/}} data, quantify the resulting actuarial unfairness across socioeconomic groups in the current annuity divisor, and propose four implementable income-dependent annuitization rules.

This question is important for two reasons. First, within China's urban employee pension scheme, workers contribute 8\% of wages to an individual account, and the accumulated notional balance is converted at retirement using a statutory divisor that depends only on retirement age. The design is therefore simple, but it also imposes a strong homogeneity assumption: retirees of the same age are treated as if they had the same post-retirement survival prospects. Second, existing research has already documented substantial inequality within China's pension system, including gender gaps, design-induced disparities, and broader fragmentation across participants and regions \citep{shen2016patterns,fang2018chinese,lu2023design,li2024gender,li2025gender}. Our focus is different but complementary. We ask whether the annuity conversion rule itself, even when applied uniformly, amplifies socioeconomic inequality because longevity differs across income groups.

This concern is closely related to the growing evidence on mortality inequality. A developing literature documents substantial differences in mortality and life expectancy across socioeconomic and area-level groups in China \citep{bai2023projections,zhu2023socioeconomic,peng2024area}. Once such differences exist, a uniform annuity divisor mechanically favors groups with lower mortality, since they expect to collect benefits for longer. This logic is well understood in the NDC literature, where heterogeneous longevity has been shown to matter for annuity design and fairness \citep{del2022fairness,jijiie2022mortality}. In the Chinese setting, the issue is particularly salient because income is directly linked both to the accumulation of the individual account and, potentially, to post-retirement survival. A reform that ignores mortality heterogeneity may therefore miss an important source of implicit redistribution. This mechanism is also connected to the broader economics literature on annuitization, beginning with the role of annuities in insuring uncertain lifetimes \citep{yaari1965uncertain}, and continuing through work on annuity welfare and household annuitization decisions \citep{brown2001private,davidoff2005annuities}.

Turning this idea into a credible empirical evaluation is not straightforward. Subgroup mortality data in China are limited, survey panels are short, and naively flexible multi-population mortality models can become unstable when estimated on sparse socioeconomic observations. At the same time, restricting attention to aggregate mortality misses the distributional issue entirely. Our strategy is therefore to combine two data sources and two levels of aggregation: we use national mortality data to identify the common period effect, and CHARLS subgroup data to identify cross-group variation in baseline mortality. We then structure those group-specific baseline schedules using Hermite splines, which provide a parsimonious yet flexible representation of adult mortality differences.\footnote{Hermite splines are classical tools in numerical analysis and computer graphics because they combine local control with flexible shape preservation; see, for example, \citet{fritsch1980monotone}, \citet{hyman1983accurate}, and \citet{kochanek1984interpolating}. In mortality modeling, they have been used much more recently; see \citet{richards2020hermite}, \citet{tang2023hermite}, \citet{huang2025towards}.}

In response to these challenges, this paper makes three contributions. First, we propose a practical framework for modeling socioeconomic mortality heterogeneity when subgroup time series are short and noisy. Our specification combines a mortality-differentiated Lee--Carter structure, closely related to the common-factor component of the Li--Lee model \citep{li2005coherent}, with a Hermite-spline representation of group-specific baseline mortality schedules. Our innovation lies in combining the common-factor decomposition with a parsimonious spline structure, normalization conditions, and shape restrictions tailored to sparse subgroup data. Empirically, we implement this framework in two steps: we estimate the common period component from national mortality data, which provide a longer and more reliable time series, and then estimate the group-specific mortality profiles from CHARLS, which provides the cross-sectional heterogeneity needed to identify income gradients in mortality. This two-step, multi-source estimation strategy allows us to model mortality heterogeneity in a setting where fully group-specific dynamic models are difficult to estimate reliably. Our use of Hermite splines is also substantively different from the recent mortality literature: rather than using them simply to fit a single-population post-retirement mortality curve, we use them to structure cross-group baseline mortality differences within a common-trend multi-group system.

Second, we apply the estimated mortality model to China's NDC pension system and quantify the actuarial unfairness embedded in the current age-only annuity divisor. Using Chinese national mortality data for 1994--2020 and CHARLS subgroup survey data for 2011--2020, we find that at retirement age 60 in the reference year 2020, the actuarially fair counting month ranges from 157.0 months in the lowest income quintile to 161.1 months in the highest, compared with the official value of 139. The implied subsidy rates, defined as the percentage by which the actuarially fair counting month exceeds the official counting month, range from 13.0\% to 15.9\% across income quintiles. At retirement age 63, the fair counting month ranges from 153.3 to 157.8 months, compared with the official value of 117, implying substantially larger subsidies of 31.0\% to 34.8\%. The monotone income gradient in both cases indicates that the current system embeds not only an overall actuarial shortfall but also a reverse transfer from poorer to richer retirees. Projections to 2040 further show that this distortion is persistent and tends to grow as mortality continues to improve.

Third, we move beyond diagnosis and evaluate four implementable income-dependent annuitization rules. The first two methods define the reform directly in terms of the average counting month, that is, the divisor applied to the entire account balance: Method 1 uses a piecewise-constant schedule by income bracket, while Method 2 uses a piecewise-linear schedule over income. The latter two methods instead define the reform in terms of the marginal counting month, that is, the divisor applied to each additional unit or account balance: Method 3 uses a piecewise-constant bracket-based marginal rule, and Method 4 uses a piecewise-linear marginal rule. Our results show that even the naive Method 1 already reduces the unfairness substantially. The more structured designs improve the policy implementation in different ways: the piecewise-linear average-divisor rule provides the closest approximation to the continuously fair benchmark, while Methods 3 and 4 impose transparent marginal conversion schedules. In particular, Methods 3 and 4 improve fairness while explicitly governing the marginal conversion of additional earnings into pension benefits, much like a marginal tax schedule governs the treatment of additional income. More broadly, the analysis shows that actuarial fairness in NDC pension design cannot be assessed without modeling heterogeneity in longevity, and that parsimonious multi-source mortality models can make such assessments feasible even when data availability is limited.

This paper relates to three strands of literature. The first studies inequality within China's pension system, with particular attention to gender inequality, regional fragmentation, and pension design \citep{shen2016patterns,fang2018chinese,lu2023design,li2024gender,li2025gender}. The second studies socioeconomic differences in mortality and life expectancy in China \citep{bai2023projections,zhu2023socioeconomic,peng2024area}. The third studies annuity design, annuitization, and the fairness implications of heterogeneous longevity. Classic work has shown that mandatory annuitization can generate substantial redistribution when mortality differs across groups \citep{brown2003redistribution,gong2008mortality}, while related pension applications study annuity design and longevity-risk management in defined contribution systems \citep{fong2011longevity,horneff2023fixed}. In the NDC context, heterogeneous longevity has been shown to matter for annuity design and fairness \citep{palmer2019annuities,del2022fairness,jijiie2022mortality,zhang2025welfare}. Closely related to our approach, \cite{huang2025towards} use individual-level linked Australian population data and Hermite-spline Poisson regression models to document substantial post-retirement mortality differentials by socioeconomic characteristics, and show that these differentials have important implications for annuity income and retirement product design. Their analysis provides complementary evidence from the Australian superannuation context, while our paper studies China’s NDC system and develops income-dependent annuitization rules within a common-trend multi-source mortality framework. In stochastic mortality modeling more broadly, the Lee--Carter model and its multi-population variants \citep{lee1992modeling,li2005coherent} remains a standard benchmark, while the Cairns--Blake--Dowd family is also widely used for adult and old-age mortality \citep{cairns2006two,hunt2021structure,dowd2022projecting}. Existing NDC fairness studies that incorporate socioeconomic mortality differentials typically rely on relatively parameter-rich multi-population mortality structures and often classify socioeconomic status by education. By contrast, we focus on income, which is directly linked to career earnings, the individual account balance, and the design of an income-dependent annuitization rule.

The remainder of the paper is organized as follows. Section~\ref{sec:hermite_model} introduces the mortality-differentiated Lee--Carter model and the Hermite-spline specification for group-specific baseline mortality. Section~\ref{sec:hermite_estimation} describes the Chinese data and the two-step estimation procedure. Section~\ref{sec:application} applies the estimated mortality model to China's NDC pension system and studies the actuarial consequences of the current annuity divisor and several progressive alternatives. Finally, Section~\ref{sec:conclusion} concludes.

\section{Modeling Mortality Differential}

\label{sec:hermite_model}

\subsection{The Mortality-Differentiated Lee-Carter Model}

Consider a population observed over ages $x \in \{x_0,\ldots,x_1\}$ and partitioned into
$J$ socioeconomic groups indexed by $j=1,\ldots,J$, where a higher value of $j$ corresponds to a higher socioeconomic group. As a benchmark, we begin with the
Lee--Carter structure:
\begin{equation}
D_{x,t} \sim \mathrm{Poisson}\!\left(E_{x,t} \cdot m_{x,t}\right),
\qquad
\log m_{x,t} = \alpha_x + \beta_x \cdot \kappa_t,
\label{eq:lc_benchmark}
\end{equation}
where $D_{x,t}$ denotes the number of deaths, $E_{x,t}$ is the exposure, $\alpha_x$ is the baseline log-mortality schedule, $\beta_x$ measures the sensitivity of age $x$ to changes in the period index, and $\kappa_t$ is the common period effect.

We use the Lee--Carter model mainly as a familiar starting point and only for illustrative purposes. For old-age mortality, the Cairns--Blake--Dowd (CBD) class and its variants are also a natural choice. The original CBD model was developed for post-age-60 mortality and uses two stochastic factors to capture changes in the level and slope of old-age mortality \citep{cairns2006two}. Subsequent work has compared CBD-type models with other stochastic mortality specifications, classified the expanding model class, extended it to socioeconomic mortality differences, and adapted it to extreme old age. See, for example, \citet{cairns2009quantitative, hunt2021structure, cairns2016modelling, dowd2022projecting}.

In a data-rich environment, one would ideally allow both group-specific age effects and group-specific period effects. Such a fully dynamic specification would accommodate the possibility that mortality improvement differs systematically across subpopulations. This is important because mortality gradients by socioeconomic status may differ not only in level, but also in their rates of change over time. In many developing-country settings, however, subgroup-specific panels are too short and noisy to support reliable estimation of separate time trends. We therefore adopt a parsimonious common-trend specification:
\begin{equation}
D_{x,j,t} \sim \mathrm{Poisson}\!\left(E_{x,j,t} \cdot m_{x,j,t}\right),
\qquad
\log m_{x,j,t} = \alpha_{x,j} + \beta_x \cdot \kappa_t,
\label{eq:mdl_lc}
\end{equation}
where $\alpha_{x,j}$ captures the baseline mortality schedule for group $j$, while $\beta_x \kappa_t$ captures the common pattern of mortality improvement over time. This specification allows mortality levels to differ across groups while borrowing the common time trend from higher-quality aggregate mortality data.

Equation~\eqref{eq:mdl_lc} is closely related to the common-factor component of the Li--Lee model \citep{li2005coherent}. In their formulation, each population has its own baseline age profile $a(x,i)$, while the time evolution is driven by a shared common factor $B(x)K(t)$; an additional population-specific term $b(x,i)k(t,i)$ is introduced only when the data support further population-level dynamics. Our specification can be interpreted as the restricted common-factor part of that framework, with $\alpha_{x,j}$ playing the role of $a(x,i)$, $\beta_x$ the role of $B(x)$, and $\kappa_t$ the role of $K(t)$, but without the group-specific period factor $b(x,i)k(t,i)$. This restriction is deliberate, since, in our setting, subgroup time series are too sparse to justify reliable estimation of additional group-specific period effects. Our contribution instead lies in imposing a structured and parsimonious representation on $\alpha_{x,j}$ through Hermite splines and in combining subgroup survey data with national mortality data in estimation.

While various parameterizations of $\alpha_{x,j}$ are possible, we use Hermite splines because they provide a low-dimensional and interpretable representation of adult mortality schedules while allowing sufficient flexibility to capture mortality convergence at advanced ages. Following \citet{richards2020hermite} and \citet{tang2023hermite}, we write
\begin{align}
    \alpha_{x,j} = \theta_j \cdot h_{00}(\tilde{x}) + \omega_j \cdot h_{01}(\tilde{x}) + \mu_{j}^{(0)} \cdot h_{10}(\tilde{x}) +  \mu_{j}^{(1)} \cdot h_{11}(\tilde{x}), \label{eq:hermite_spline}
\end{align}
where $\tilde{x} = (x - x_0)/(x_1 - x_0)$ denotes the standardized age, and $h_{00}$, $h_{01}$, $h_{10}$, and $h_{11}$ are Hermite basis functions:
\begin{align*}
    \begin{cases}
        h_{00}(\tilde{x}) = (1+2\tilde{x})(1-\tilde{x})^2\\
        h_{01}(\tilde{x}) = \tilde{x}^2 (3-2\tilde{x})\\
        h_{10}(\tilde{x}) = \tilde{x}(1-\tilde{x})^2\\
        h_{11}(\tilde{x}) = \tilde{x}^2(\tilde{x}-1).
    \end{cases}
\end{align*}
The coefficients $\theta_j$, $\omega_j$, $\mu_{j}^{(0)}$, and $\mu_{j}^{(1)}$ control the shape of the Hermite spline for group $j$.

Figure \ref{fig:hermite_hs} illustrates the Hermite basis functions, each of which carries an intuitive interpretation. The four-parameter HS structure offers sufficient flexibility to model adult mortality patterns while preserving clear interpretability. Specifically, $\alpha_{x_0,j} = \theta_j$ and $\alpha_{x_1,j} = \omega_j$, indicating that $\theta_j$ and $\omega_j$ correspond to the log mortality levels at the youngest and oldest ages, respectively, for group $j$. Furthermore, the endpoint slope parameters \(\mu^{(0)}_j\) and \(\mu^{(1)}_j\) determine the rates of increase in log mortality with respect to standardized age at \(x_0\) and \(x_1\), respectively.

\begin{figure}[!ht]
\centering
\includegraphics[width=0.75\linewidth]{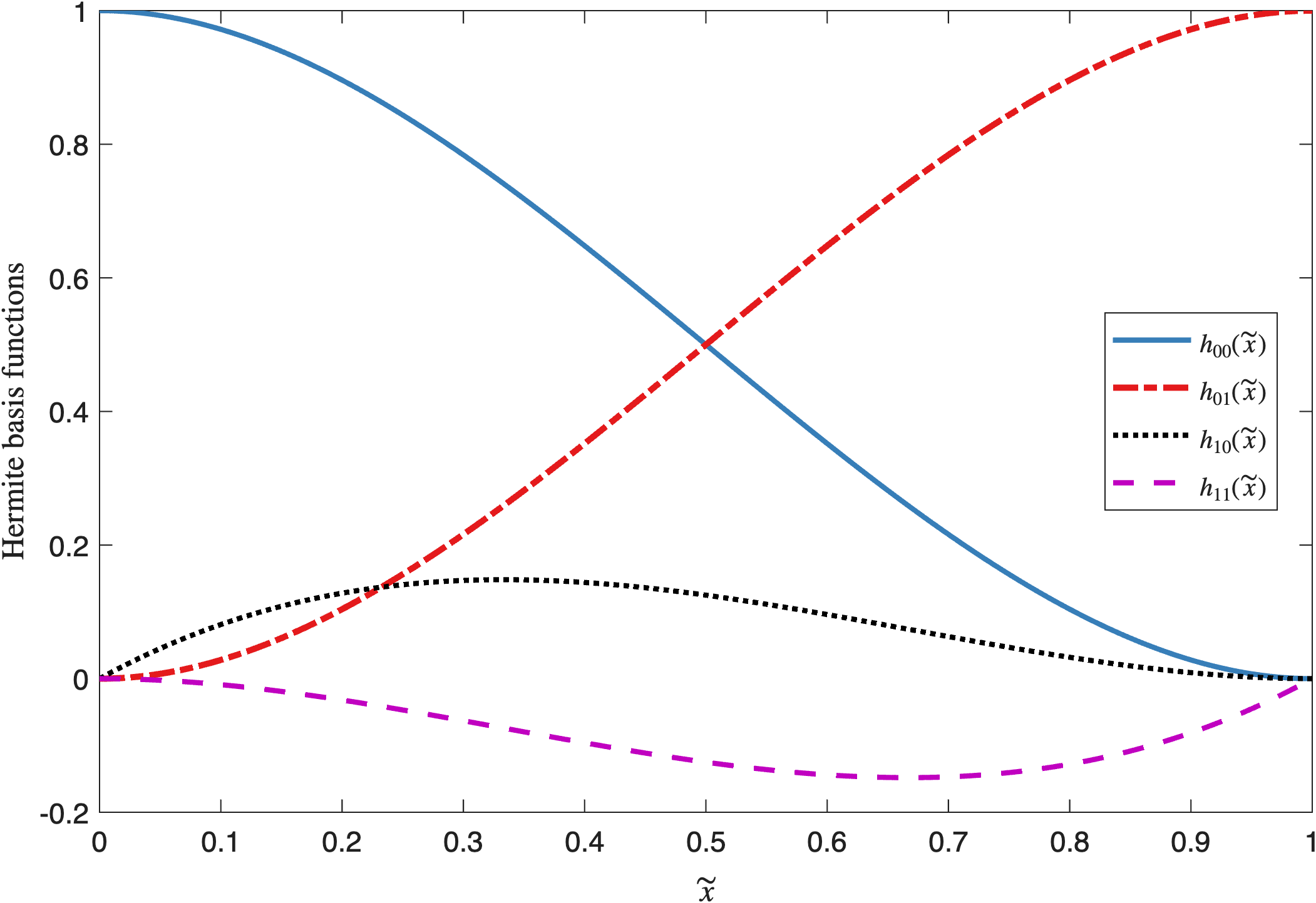}
\caption{Basis functions of Hermite splines over standardized age $\tilde{x}$.}
\label{fig:hermite_hs}
\end{figure}

\subsection{Comparison with Gompertz}

The Hermite spline model generalizes the Gompertz model. By imposing the constraint $\mu_{j}^{(0)} = \mu_{j}^{(1)} = \omega_j - \theta_j$, i.e., fixing the slope of a straight line connecting log mortality at the youngest and oldest ages, the Hermite spline reduces to a two-parameter Gompertz form:
\begin{align*}
    \alpha_{x,j} = a_j + b_j \cdot x,
\end{align*}
where $a_j$ is the log mortality level at age $0$ and $b_j$ denotes the log mortality growth rate for group $j$.

The left panel of Figure \ref{fig:hermite_gompertz} displays the estimates of $ \alpha_{x,j}$ from the Gompertz model, fitted to the mortality rates calculated from the CHARLS survey data (see Section~\ref{sec:hermite_estimation} for mortality rate construction). A key shortcoming becomes immediately apparent: the model implies crossover behavior between income groups, e.g., individuals in the higher-income groups exhibit lower mortality at younger ages but higher mortality at older ages compared to lower-income groups. This outcome contradicts both empirical findings and intuition and stems directly from the linear log-mortality assumption in the Gompertz model.

\begin{figure}[!ht]
    \centering
    \begin{subfigure}{0.49\linewidth}
        \centering
        \includegraphics[width=\linewidth]{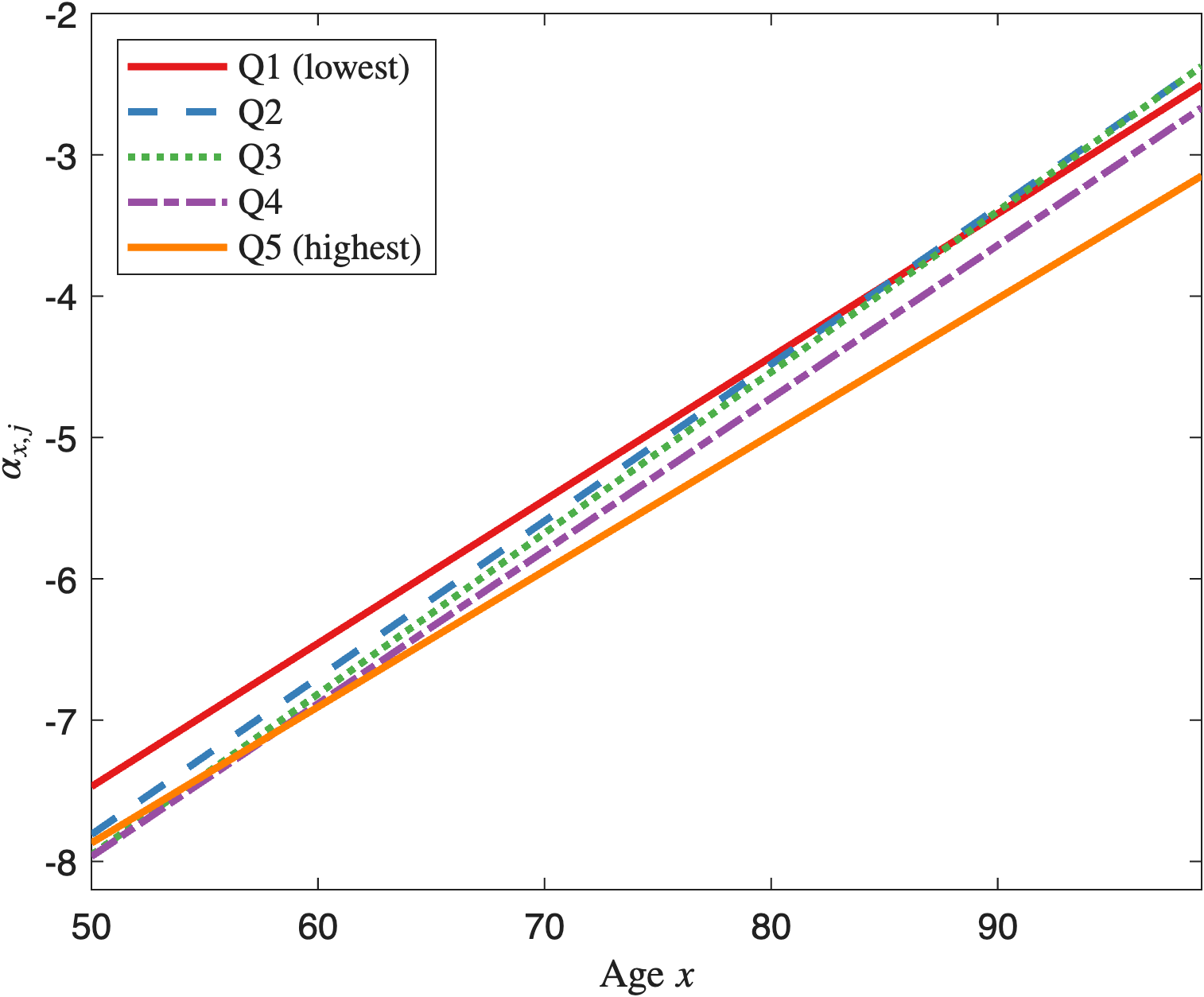}
        \caption{Unconstrained}
    \end{subfigure}
    \hfill
    \begin{subfigure}{0.49\linewidth}
        \centering
        \includegraphics[width=\linewidth]{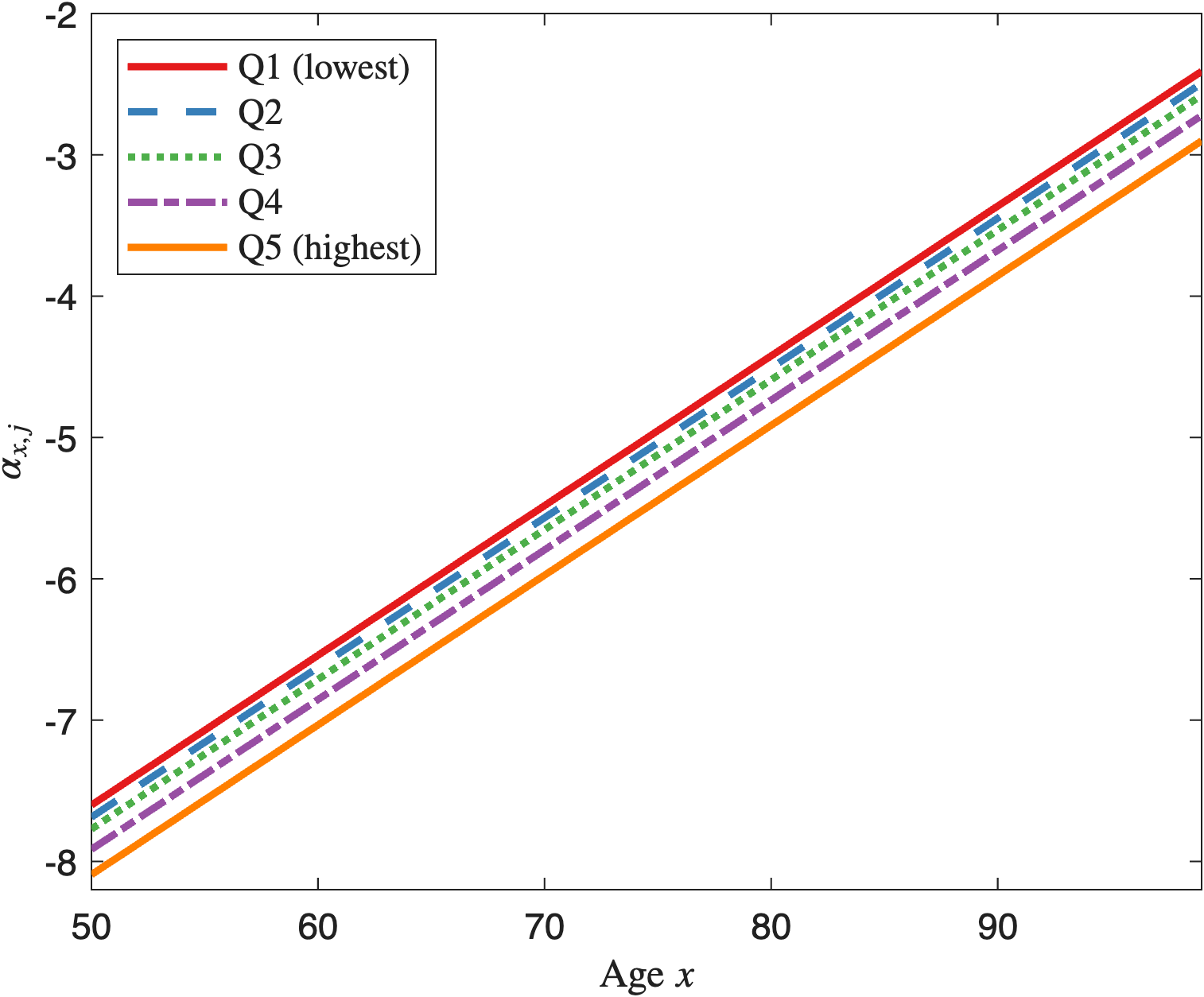}
        \caption{Constrained}
    \end{subfigure}
    \caption{Log mortality curves of the fitted Gompertz models by income quintile.}
    \label{fig:hermite_gompertz}
\end{figure}

To eliminate such crossovers, one might impose monotonic constraints on the coefficients: (i) $a_j \leq a_i$ for $j > i$, and (ii) $b_j \leq b_i$ for $j > i$. These constraints ensure that higher-income groups start with lower mortality at age $x_0$ and maintain a mortality advantage throughout the age range. The estimates of $ \alpha_{x,j}$ from the constrained Gompertz model are shown in the right panel of Figure \ref{fig:hermite_gompertz}. Under these constraints, the model effectively forces the slopes $b_j$ to be identical across groups, producing parallel log-mortality trajectories, which will in turn lead to a substantial deterioration in the model's goodness-of-fit.

More fundamentally, the Gompertz model --- whether constrained or not --- fails to capture two key features observed in human mortality: (i) the \textit{compensation law of mortality}, which refers to mortality convergence over age, whereby higher initial mortality levels (i.e., $a_i$) are often accompanied by lower mortality growth rates (i.e., $b_i$); and (ii) the \textit{mortality plateau}, where the rate of mortality increase slows at advanced ages.\footnote{See, among others, \cite{strehler1960general,gavrilov2005reliability} for compensation law of mortality, and \cite{gavrilov2005reliability} and \cite{weitz2001explaining} for mortality plateau.} 

The Hermite spline model can flexibly accommodate both phenomena. The mortality plateau is captured by setting the slope near the oldest age, $\mu_{j}^{(1)}$, close to zero. Mortality convergence can be achieved through appropriate differences in intercepts $\theta_j$ across groups. As shown in \cite{richards2020hermite}, even when groups differ only in $\theta_j$, Hermite splines can ensure eventual convergence at older ages without inducing crossover behavior.

\subsection{Hermite Spline Model for Mortality Heterogeneity}

To extend the Hermite spline specification to socioeconomic mortality heterogeneity, we must decide which coefficients are group-specific and which are shared across groups. In the Hermit spline specification~\eqref{eq:hermite_spline}, $\omega_j$ determines the log mortality level at the limiting age $x_1$, $\mu_{j}^{(0)}$ determines the mortality gradient at the youngest fitted age $x_0$, and $\mu_{j}^{(1)}$ determines the mortality gradient at the limiting age $x_1$. In general, these parameters can in principle all be specified as either common across groups or group-specific, depending on the application and the degree of heterogeneity one wishes to allow. This flexibility makes the Hermite spline framework well suited to multi-group mortality modeling. In our analysis, we focus on four grouped Hermite spline specifications. In all of them, we keep the terminal slope $\mu^{(1)}$ common across groups and vary the restrictions on the terminal level $\omega_j$ and the initial slope $\mu_{j}^{(0)}$.

\begin{table}[!ht]
    \centering
    \caption{Parameter settings of the Hermite spline model for capturing socioeconomic mortality heterogeneity.}
    \begin{tabular}{c|c|c}
    \toprule
       Model  & Group-Specific Parameters & Group-Shared Parameters  \\
    \midrule
       HSM-I   & $\theta_{j}$, $\omega_{j}$, $\mu_{j}^{(0)}$ & $\mu^{(1)}$ \\
       HSM-II  & $\theta_{j}$, $\omega_{j}$ & $\mu^{(0)}$, $\mu^{(1)}$   \\
       HSM-III & $\theta_{j}$, $\mu_{j}^{(0)}$ & $\omega$, $\mu^{(1)}$ \\
       HSM-IV  & $\theta_{j}$ & $\omega$, $\mu^{(0)}$, $\mu^{(1)}$  \\
    \bottomrule
    \end{tabular}
    \label{tab:hermite_family_group}
\end{table}

These four specifications differ in how much cross-group structure is imposed on the baseline mortality curves. HSM-I is the least restrictive specification: apart from the common terminal slope $\mu^{(1)}$, all remaining parameters are allowed to differ across groups. HSM-II additionally imposes $\mu_{j}^{(0)}=\mu^{(0)}$, so mortality gradients are also required to be the same at the youngest fitted age $x_0$. HSM-III instead imposes $\omega_j=\omega$, so all groups share the same mortality level at the limiting age $x_1$ while retaining group-specific mortality gradients at $x_0$. HSM-IV is the most restrictive of the four, combining both restrictions so that groups share the same mortality level at $x_1$ and the same mortality gradient at $x_0$.

In the applications below, we illustrate the modeling performance of our proposed model using HSM-III.\footnote{In an auxiliary cross-sectional validation exercise using U.S. mortality data for 1982--2019, HSM-III also provides the best balance between fit and parsimony among the HSM specifications considered here, as measured by AIC and BIC.} The restriction $\omega_j=\omega$ implies that the fitted mortality levels of all groups converge at the limiting age $x_1$, which is biologically reasonable for old-age mortality. At the same time, HSM-III leaves $\mu_{j}^{(0)}$ group-specific, so the model does not force all groups to have the same mortality gradient at the youngest fitted age. By contrast, specifications that impose $\mu_{j}^{(0)}=\mu^{(0)}$ require a common initial slope across groups, which seems unnecessarily restrictive in our setting. We also estimated the proposed model using other HSM specifications, and the main qualitative conclusions are similar, so we do not report the full set of results.

A sufficient non-crossover condition for HSM-III, when groups are ordered from low to high socioeconomic status, is\footnote{Given $\theta_j<\theta_i$ for $j>i$, then the non-crossover condition is satisfied if for all $0<\tilde{x}<1$, 
\begin{align*}
    (\theta_j-\theta_i)\cdot h_{00}(\tilde{x}) + (\mu_{j}^{(0)} - \mu_{i}^{(0)})\cdot h_{10}(\tilde{x}) < 0 \implies  &(\mu_{j}^{(0)} - \mu_{i}^{(0)}) <  \frac{1+2\tilde{x}}{\tilde{x}}(\underset{>0}{\underbrace{\theta_i-\theta_j}})\implies (\mu_{j}^{(0)} - \mu_{i}^{(0)}) < \underset{=3}{\underbrace{\inf_{\tilde{x}\in(0,1)}  \left( \frac{1+2\tilde{x}}{\tilde{x}}\right)}} (\theta_i-\theta_j).
\end{align*}

}
\begin{equation}
\mu_{j}^{(0)}-\mu_{i}^{(0)}\le -3(\theta_j-\theta_i), \qquad j>i,
\end{equation}
which can either be checked after estimation or imposed directly as a linear inequality constraint during estimation. The resulting HSM-III specification is therefore
\begin{align*}
    \alpha_{x,j} = \theta_j\cdot h_{00}(\tilde{x}) + \omega \cdot h_{01}(\tilde{x}) + \mu_{j}^{(0)}\cdot h_{10}(\tilde{x}) + \mu^{(1)}\cdot h_{11}(\tilde{x}).
\end{align*}

\section{Data and Estimation}
\label{sec:hermite_estimation}

The mortality-differentiated Lee--Carter model introduced in Section~\ref{sec:hermite_model} is estimated using two Chinese data sources. First, the aggregate mortality trend $\kappa_t$ and the age-sensitivity profile $\beta_x$ are estimated from Chinese national uni-sex mortality data for 1994--2020 obtained from the \emph{National Population Census Yearbooks} and the \emph{National 1\% Sample Survey Yearbooks} issued by the Bureaus of Statistics of China.\footnote{For a more detailed description of the dataset, see \cite{li2019bayesian} and \cite{li2026national}.} Second, conditional on $(\kappa_t,\beta_x)$, the group-specific baseline effects $\alpha_{x,j}$ are estimated from uni-sex mortality data obtained from the China Health and Retirement Longitudinal Study (CHARLS). This two-step design allows us to exploit the broader coverage and longer time horizon of the national data for the common time component, while using the CHARLS data to identify cross-sectional mortality heterogeneity by income quintiles.

CHARLS is a nationally representative survey administered by the research institute at Peking University. It covers over 150 counties and has been conducted in the years 2011, 2013, 2015, 2018, and 2020. It is chosen for two main reasons: (1) it includes rich questionnaire content, such as income, health status, and wealth, which makes it more comprehensive than other available surveys or databases; and (2) it targets individuals aged 45 and older, with the majority near or past retirement age, making it particularly suitable for pension-related analysis. However, as a survey-based dataset, it requires preprocessing and presents certain limitations. In particular:  
(1) mortality is only recorded in terms of whether a respondent died between two waves, without exact death dates;  
(2) although CHARLS is one of the largest datasets of its kind in China, it is still limited in scope, containing 17,705 respondents in 2011, and only 17,678 records after preliminary cleaning (e.g., removing missing values); and
(3) self-reported income and wealth data may contain measurement errors.

We process the CHARLS data as follows:  
(1) we restrict the sample to individuals aged 50 and above;\footnote{Only about 0.2\% of the respondents are below 50.}  
(2) we calculate each respondent's average annual income across all waves in which the respondent is observed and use it as the criterion for assigning respondents to the five income quintiles; in order to reduce the influence of outliers we trim the average income by excluding the top and bottom 2.5\%;\footnote{Empirical studies often trim or winsorize the top and bottom 0.5\% or 1\% of the income distribution to reduce the influence of outliers. In the CHARLS data, however, 1.4\% of respondents reported a zero income in all waves, while a few respondents reported unrealistically high annual incomes, e.g., above 600 million CNY. Such extreme values are common in survey-based income data and likely reflect reporting frictions or measurement error \citep{moore2000income}. In the baseline specification, we therefore exclude the top and bottom 2.5\% of the sample. As robustness checks, we also consider three alternative preprocessing rules: excluding the top and bottom 1.4\% of the sample, excluding the top and bottom 2\%, and retaining the full sample. The qualitative results are similar across these alternatives, so we do not report them here. Results from these alternative preprocessing choices are available upon request.} (3) we divide the remaining sample into five quintiles based on income adjusted to 2020 currency values. Table~\ref{tab:hermite_quintiles} reports the mean income and range for each quintile.

\renewcommand{\arraystretch}{1.2}
\begin{table}[!ht]
	\centering  
    \caption{Income-based socioeconomic quintiles in the CHARLS dataset, income adjusted to 2020 (CNY).} 
    \label{Tab.3}
	\begin{tabular}{ccc}  
		\toprule  
        Group & Mean Income  &  Income Range \\
		\midrule  
	    Quintile 1 & 2,181  & $[568,\ 3,847]$           \\
       	Quintile 2 & 6,131  & $(3,847,\ 8,838]$          \\
        Quintile 3 & 12,902 & $(8,839,\ 17,651]$     \\
        Quintile 4 & 23,897 & $(17,652,\ 31,300]$  \\
        Quintile 5 & 51,599 & $(31,302,\ 122,534]$ \\
		\bottomrule  
	\end{tabular}     
    \label{tab:hermite_quintiles}
\end{table}  

Figure~\ref{fig:charls-overview} provides a compact overview of the estimation sample constructed from the five CHARLS waves 2011, 2013, 2015, 2018, and 2020. The top-left panel reports person-year exposure by age and income quintile, while the top-right panel reports annualized death counts. The bottom-left panel shows pooled log central death rates by age and income quintile, and the bottom-right panel displays the distribution of career-average annual income on a log scale. Two features are especially relevant for the estimation below. First, exposure is concentrated between ages 50 and 80 and declines rapidly at advanced ages, which favors a parsimonious mortality specification. Second, there is a clear income gradient in observed mortality, especially between the lowest-income quintile and the highest-income quintile. The income histogram also makes transparent the concentration of observations in the lower part of the income distribution.

\begin{figure}[!ht]
    \centering
    \begin{subfigure}{0.49\textwidth}
        \centering
        \includegraphics[width=\linewidth]{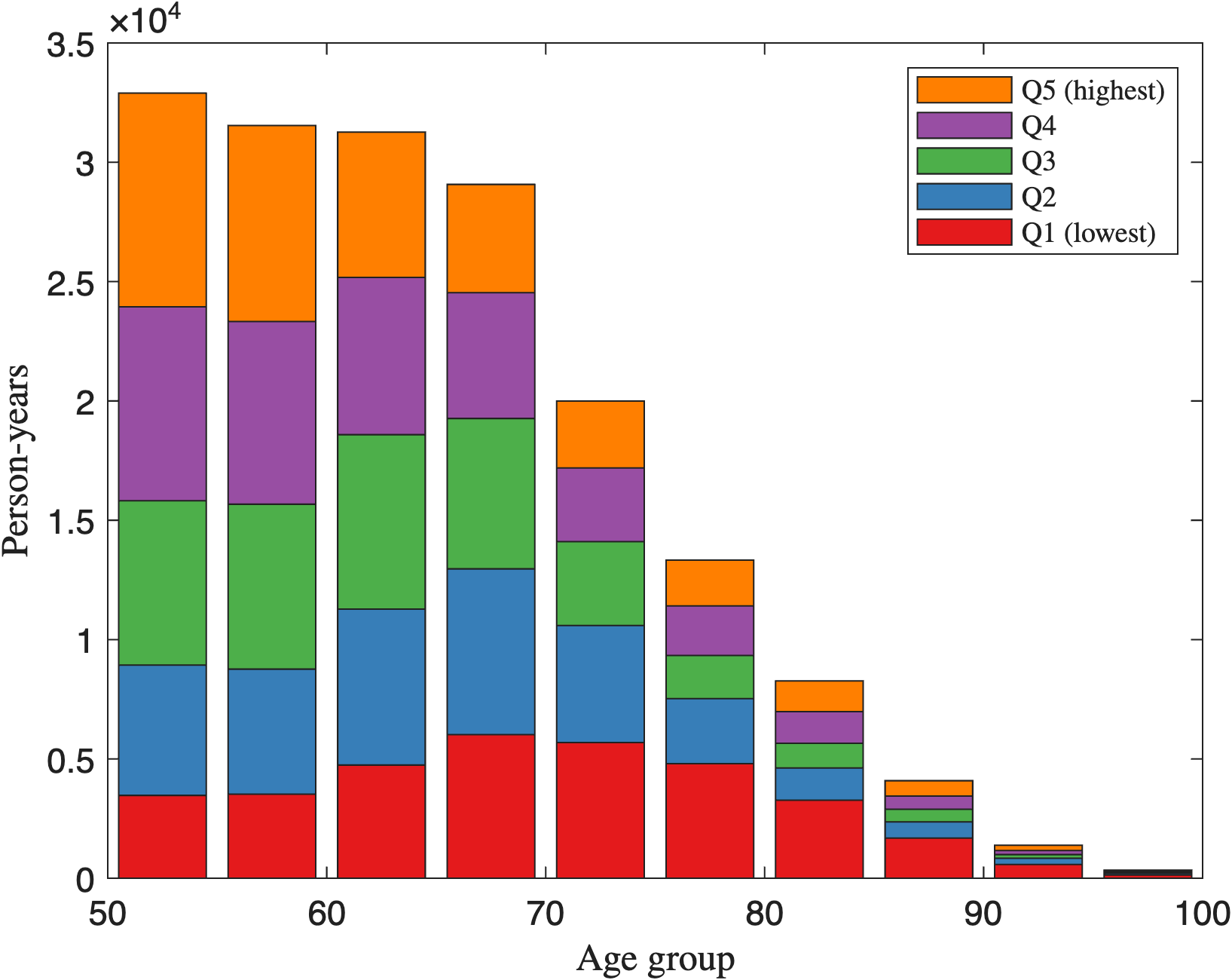}
        \caption{Exposure count}
    \end{subfigure}
    \hfill
    \begin{subfigure}{0.49\textwidth}
        \centering
        \includegraphics[width=\linewidth]{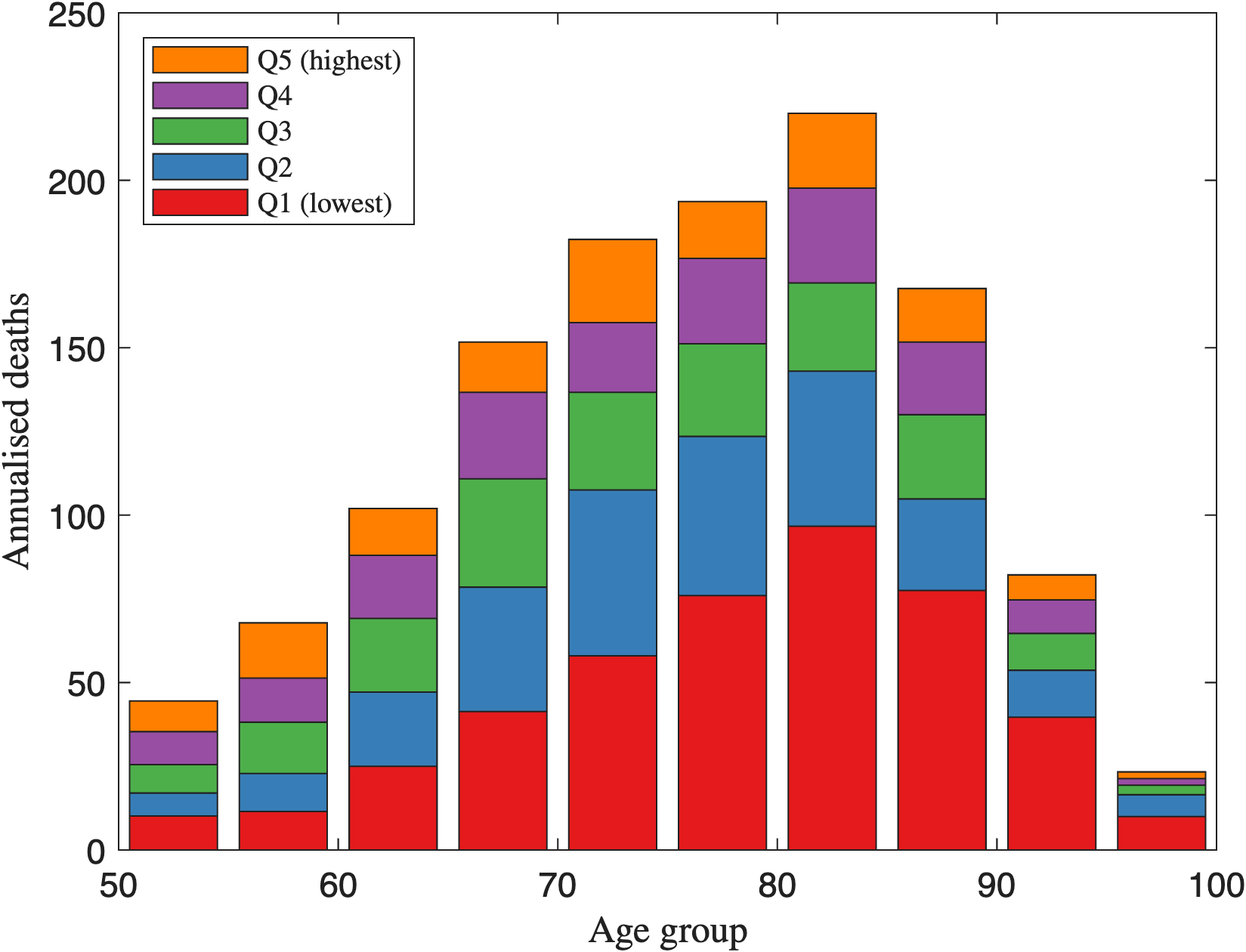}
        \caption{Death count}
    \end{subfigure}

    \vspace{0.5em}

    \begin{subfigure}{0.49\textwidth}
        \centering
        \includegraphics[width=\linewidth]{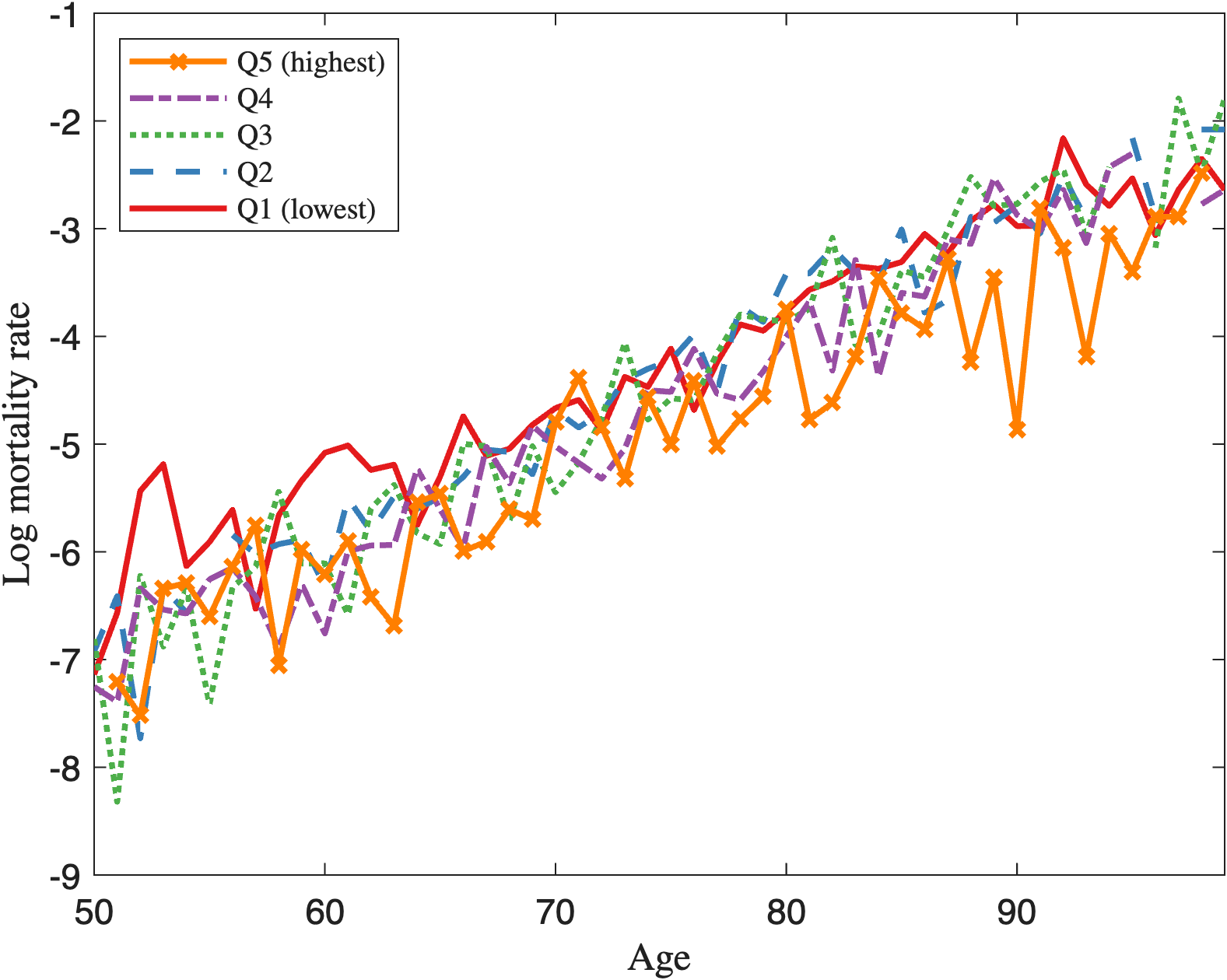}
        \caption{Log mortality rate}
    \end{subfigure}
    \hfill
    \begin{subfigure}{0.49\textwidth}
        \centering
        \includegraphics[width=\linewidth]{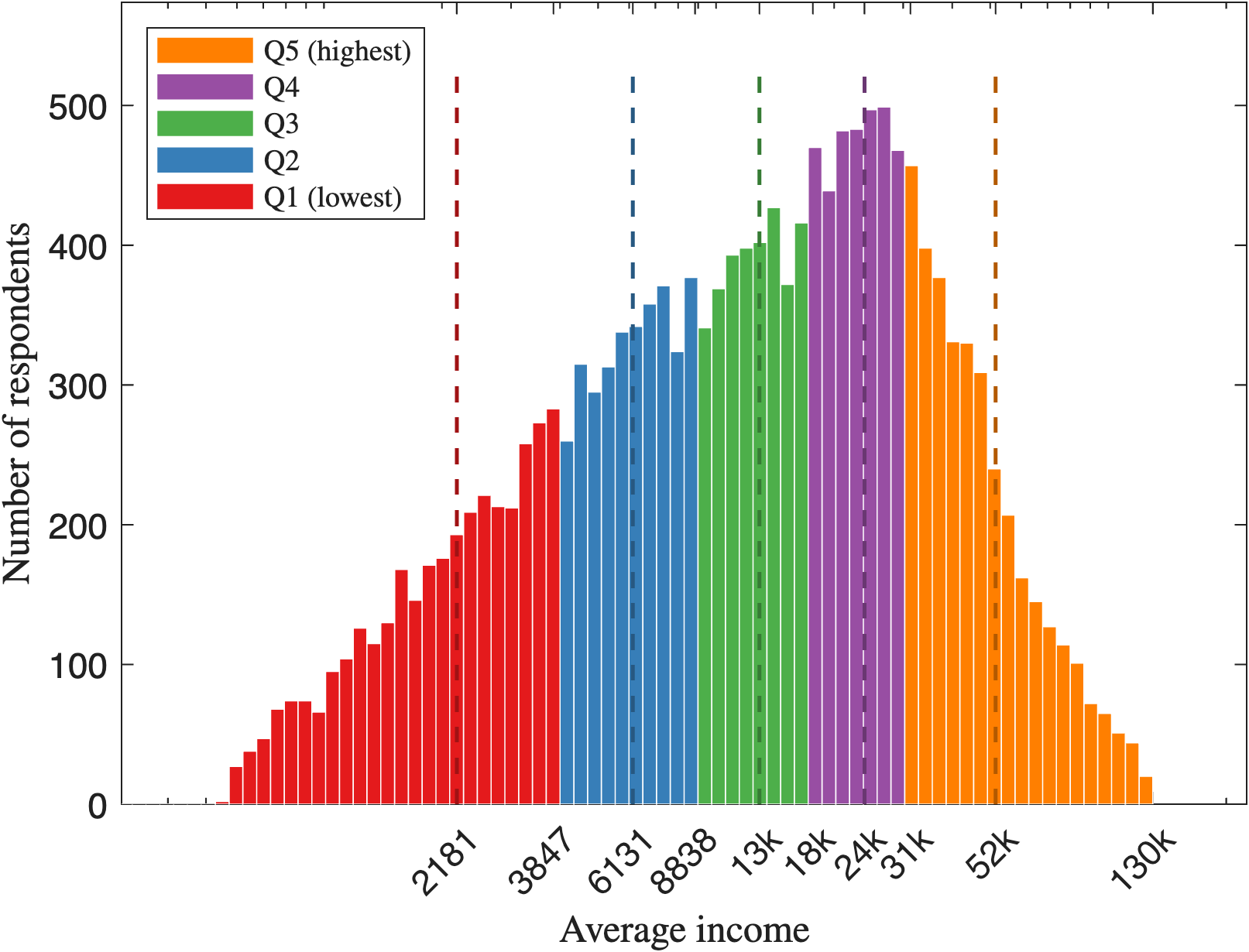}
        \caption{Average income}
    \end{subfigure}
    \caption{Summary of the CHARLS dataset used for estimation.}
    \label{fig:charls-overview}
\end{figure}

The estimation is implemented in two steps. In Step 1, we fit a standard Lee--Carter model to national mortality data in order to estimate a common age-sensitivity profile and a common time trend. In Step 2, we use the estimated national time component to remove calendar-time variation from the CHARLS data, collapse the panel into an LC-adjusted cross section, and then estimate the group-specific baseline mortality schedules. In this way, the national data provide a much longer and cleaner time series for estimating the common mortality trend, whereas the CHARLS provides the cross-sectional heterogeneity needed to identify mortality differences across income groups.

For the national data, we estimate
\begin{align}
D^{\mathrm{nat}}_{x,t}
    &\sim \mathrm{Poisson}\!\left(
        E^{\mathrm{nat}}_{x,t}
        \exp\!\left(\alpha_x + \beta_x \kappa_t\right)
    \right),
    \label{eq:nat_lc}
\end{align}
subject to the identifying restrictions
\begin{align*}
\sum_x \beta_x = 1,
\qquad
\kappa_{2020}=0.
\end{align*}
The first restriction resolves the scale indeterminacy in the Lee--Carter factorization, while the second anchors the baseline mortality schedule to the reference year 2020. The national Lee--Carter model in Equation~\eqref{eq:nat_lc} is fitted using Poisson maximum-likelihood estimation.

\begin{figure}[!ht]
    \centering
    \begin{subfigure}{0.49\textwidth}
        \centering
        \includegraphics[width=\linewidth]{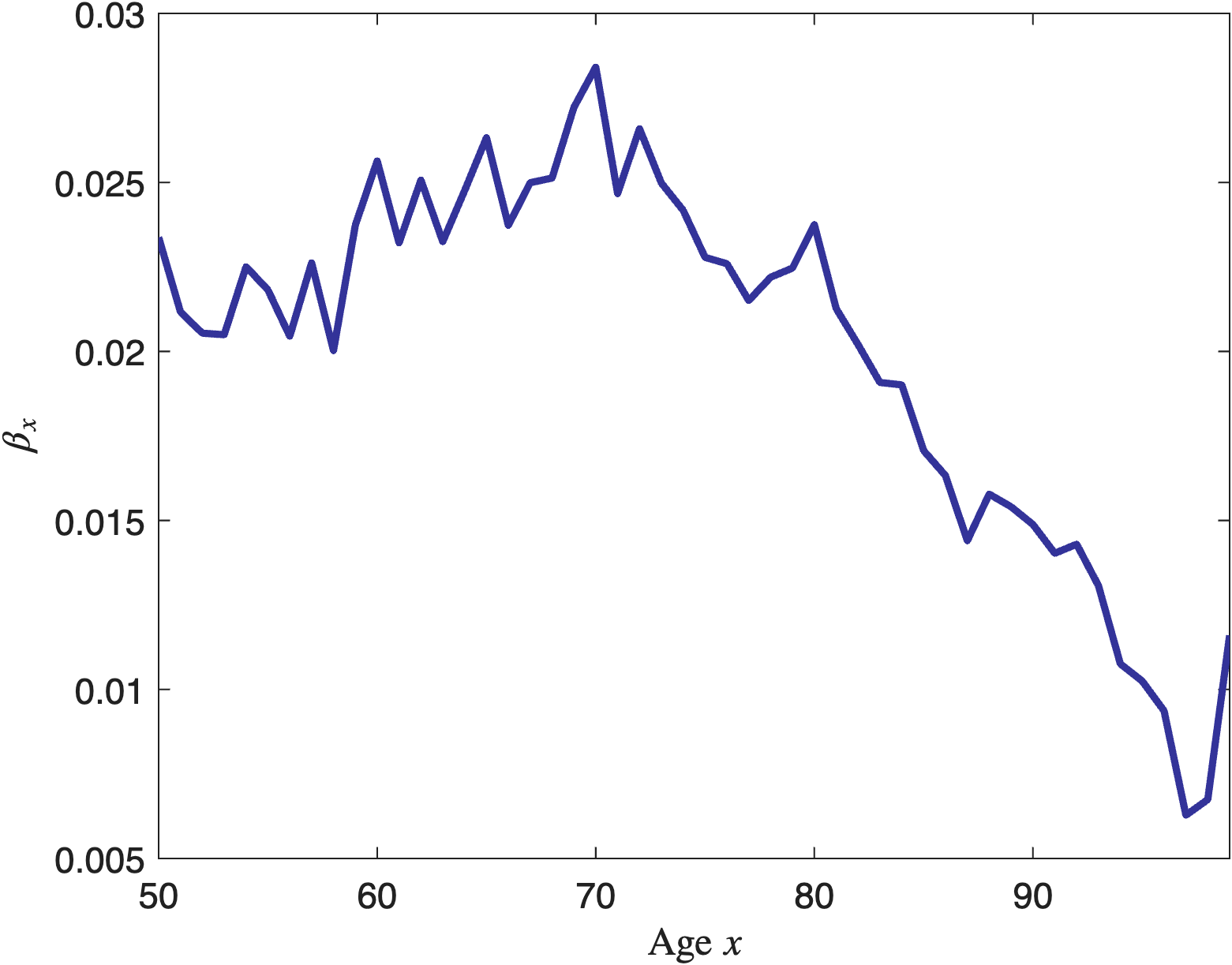}
        \caption{Age-sensitivity parameters}
    \end{subfigure}
    \hfill
    \begin{subfigure}{0.49\textwidth}
        \centering
        \includegraphics[width=\linewidth]{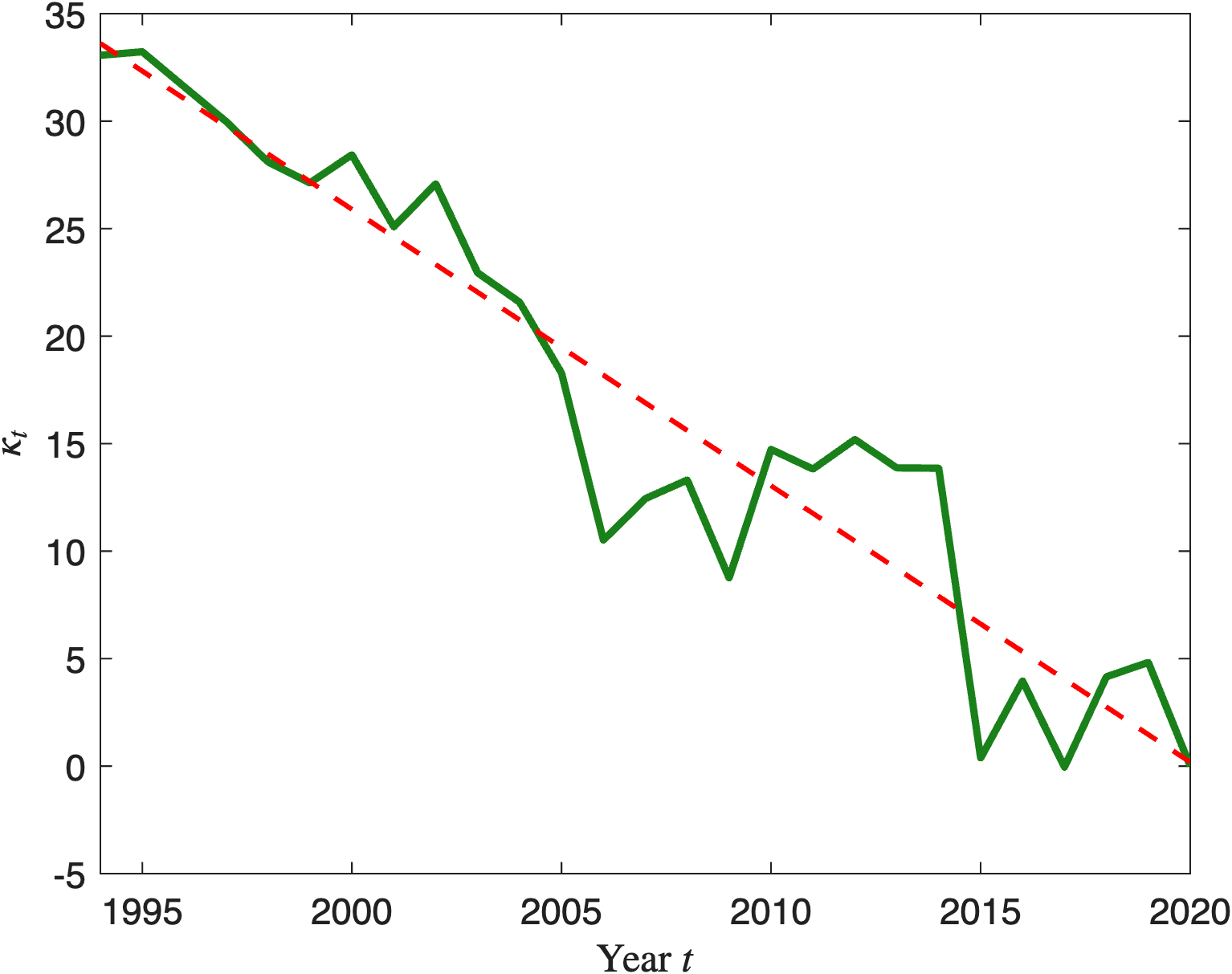}
        \caption{Time trend parameters}
    \end{subfigure}
    \caption{Lee--Carter parameters fitted to China's national mortality data.}
    \label{fig:lc-components}
\end{figure}

Figure~\ref{fig:lc-components} reports the estimated national Lee--Carter components. The age-sensitivity profile $\widehat{\beta}_x$ is positive over the fitted age range and varies non-uniformly by age, while the time index $\widehat{\kappa}_t$ shows the long-run decline in aggregate mortality over 1994--2020. The normalization $\widehat{\kappa}_{2020}=0$ anchors the baseline mortality schedule to the 2020 reference year used in the subsequent CHARLS-based estimation.

For the CHARLS data, the estimation begins from the uni-sex mortality data constructed by age, income quintile, and survey wave. Let \(\mathcal{Y}=\{2011,2013,2015,2018\}\) denote the starting waves of the four observed inter-wave mortality intervals in the five-wave CHARLS sample, and let \(\Delta_y\) denote the length of the interval until the next wave:
\begin{align*}
\Delta_{2011}=2,
\qquad
\Delta_{2013}=2,
\qquad
\Delta_{2015}=3,
\qquad
\Delta_{2018}=2.
\end{align*}
For each age $x$, income quintile $j$, and wave $y$, we let
\begin{align*}
E_{x,j,y} &= l_{x,j,y}, \qquad D_{x,j,y} = \frac{d_{x,j,y}}{\Delta_y},
\end{align*}
where $l_{x,j,y}$ is the group-specific exposure proxy and $d_{x,j,y}$ is the observed number of deaths over the full inter-wave interval. Thus, $D_{x,j,y}$ is an annualized death count, and
\begin{align*}
m^{\mathrm{obs}}_{x,j,y} = \frac{D_{x,j,y}}{E_{x,j,y}}
\end{align*}
is the corresponding annualized central death-rate proxy.

Under the mortality-differentiated Lee--Carter specification,
\begin{align*}
\ln m_{x,j,y} = \alpha_{x,j} + \beta_x \kappa_y,
\end{align*}
so that
\begin{align*}
m_{x,j,y} = \exp(\alpha_{x,j}) \exp(\beta_x \kappa_y).
\end{align*}
After obtaining $\widehat{\beta}_x$ and $\widehat{\kappa}_y$ from Step 1, we then construct an LC-adjusted pooled sample for each age-group cell:
\begin{align*}
D^{\mathrm{pool}}_{x,j}
    &= \sum_{y \in \mathcal{Y}} D_{x,j,y}, \\
E^{\mathrm{eff}}_{x,j}
    &= \sum_{y \in \mathcal{Y}}
       E_{x,j,y}\exp\!\left(\widehat{\beta}_x \widehat{\kappa}_y\right).
\end{align*}
This construction follows from the model-implied Poisson mean,
\begin{align*}
\mathbb{E}\!\left[D^{\mathrm{pool}}_{x,j}\right]
    =
    \exp(\alpha_{x,j})
    \sum_{y \in \mathcal{Y}}
    E_{x,j,y}\exp\!\left(\beta_x \kappa_y\right),
\end{align*}
which shows that, conditional on $(\beta_x,\kappa_y)$, the second step can be written as a static cross-sectional estimation problem for $\alpha_{x,j}$.

After constructing the sample, we then fit the HSM-III specification to the pooled pair
$\left(D^{\mathrm{pool}}_{x,j}, E^{\mathrm{eff}}_{x,j}\right)$. The objective function is
\begin{align}
Q(p)
=
\sum_{x,j}
\left[
D^{\mathrm{pool}}_{x,j}\,\alpha_{x,j}(p)
-
E^{\mathrm{eff}}_{x,j}\exp\!\left(\alpha_{x,j}(p)\right)
\right],
\label{eq:step2_obj}
\end{align}
where $p$ is the parameter vector. Under the HSM-III specification, we have 
\begin{align*}
\alpha_{x,j}
=
\theta_j h_{00}(\tilde{x})
+
\omega h_{01}(\tilde{x})
+
\mu_{j}^{(0)} h_{10}(\tilde{x})
+
\mu^{(1)} h_{11}(\tilde{x}),
\qquad
\tilde{x} = \frac{x-x_0}{x_1-x_0},
\end{align*}
with $x_0=50$ and $x_1=120$. Thus, $\theta_j$ and $\mu_{j}^{(0)}$ are group-specific, while $\omega$ and $\mu^{(1)}$ are shared across groups.

The baseline HSM-III estimation imposes the following shape restrictions:
\begin{align*}
\theta_{j+1} &\le \theta_j, && j=1,\ldots,J-1, \\
0 \le \mu_{j}^{(0)} &\le \mu_{j+1}^{(0)}, && j=1,\ldots,J-1, \\
(\mu_{j+1}^{(0)}-\mu_{j}^{(0)}) & \le - 3(\theta_{j+1}-\theta_j),
&& j=1,\ldots,J-1.
\end{align*}
These restrictions ensure that richer groups start from lower mortality at age $x_0$, have weakly steeper initial slopes, and satisfy a sufficient non-crossover condition over the fitted age range.

After the model is estimated, the full fitted mortality surface is then reconstructed as
\begin{align*}
\widehat{m}_{x,j,y}
=
\exp\!\left(
\widehat{\alpha}_{x,j}
+
\widehat{\beta}_x \widehat{\kappa}_y
\right).
\end{align*}
Because the national Lee--Carter component is normalized by $\widehat{\kappa}_{2020}=0$, the estimated $\widehat{\alpha}_{x,j}$ should be interpreted as the group-specific log-mortality schedule at the 2020 reference year.

In the baseline estimation, we set $x_1=120$ as the limiting age. Figure~\ref{fig:hsm-baseline} reports the estimated HSM-III baseline curves $\widehat{\alpha}_{x,j}$ by income quintile. The fitted curves are strictly ordered by income, with lower-income groups exhibiting higher baseline mortality, and they become closer at advanced ages. 

\begin{figure}[!ht]
    \centering
    \includegraphics[width=0.7\textwidth]{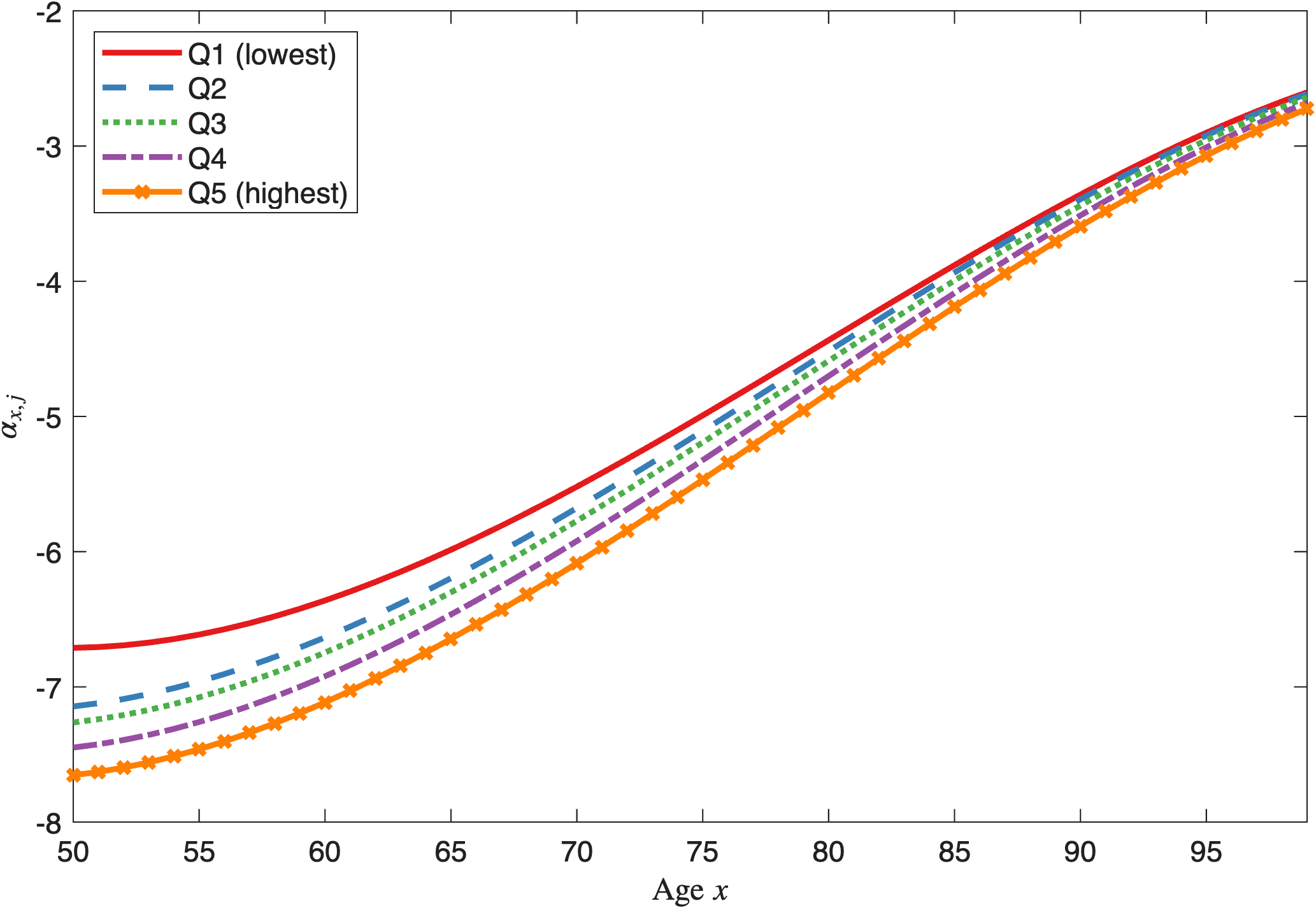}
    \caption{Log mortality curves of the fitted HSM-III baseline model by income quintile.}
    \label{fig:hsm-baseline}
\end{figure}

\section{Application to China's NDC Pension System}
\label{sec:application}

This section applies the estimated mortality model to China's notional defined contribution (NDC) pension system. Throughout, we focus on the pooled-sex results and use the period life table unless stated otherwise. Under the current rule, the monthly pension is obtained by dividing the individual's notional account balance by an official age-specific counting month. Since the official counting month depends only on retirement age, it cannot be actuarially fair for all income groups once mortality differs systematically by income.

\subsection{Actuarial unfairness under the official rule}
\label{subsec:official_unfairness}

Let $M_x^{\mathrm{off}}$ denote the official counting month for retirement age $x$. For an individual in income quintile $j$ who retires at age $x$ in calendar year $t$, let $\ddot{a}^{(12)}_{x,j,t}$ denote the actuarial present value of a life annuity paying one unit per year in monthly instalments, valued using the mortality schedule estimated in Section~\ref{sec:hermite_estimation}. The corresponding actuarially fair counting month is
\begin{equation*}
M^{\mathrm{fair}}_{x,j,t} := 12\,\ddot{a}^{(12)}_{x,j,t}.
\end{equation*}
Under the official rule, the proportional actuarial subsidy received by individuals in quintile $j$ is
\begin{equation}
\Lambda_{x,j,t}
:=
\frac{M^{\mathrm{fair}}_{x,j,t}}{M_x^{\mathrm{off}}} - 1
=
\frac{12\,\ddot{a}^{(12)}_{x,j,t}}{M_x^{\mathrm{off}}} - 1.
\label{eq:lambda_official}
\end{equation}
A positive value of $\Lambda_{x,j,t}$ means that the official divisor is too low for that group, so the retiree receives a subsidy relative to actuarial fairness. A larger value of $\Lambda_{x,j,t}$ therefore indicates a larger implicit transfer from the pension system to that group.

\begin{table}[!ht]
\centering
\caption{Actuarially fair counting month $M^{\mathrm{fair}}_{x,j,t}$ and subsidy rate $\Lambda_{x,j,t}$ by income quintile. All calculations use reference year $t=2020$ ($\kappa_{2020}=0$), discount rate $r=7\%$, and period life table.}
\begin{tabular}{l c c c c c c}
\toprule
 & \multicolumn{3}{c}{Age 60} & \multicolumn{3}{c}{Age 63} \\
\cmidrule(lr){2-4}\cmidrule(lr){5-7}
Quintile & $M^{\mathrm{fair}}_{60,j,2020}$ & $M^{\mathrm{off}}_{60}$ & $\Lambda_{60,j,2020}$ & $M^{\mathrm{fair}}_{63,j,2020}$ & $M^{\mathrm{off}}_{63}$ & $\Lambda_{63,j,2020}$ \\
\midrule
Q1 (lowest)  & 157.0 & 139 & 13.0 & 153.3 & 117 & 31.0 \\
Q2           & 158.1 & 139 & 13.8 & 154.5 & 117 & 32.0 \\
Q3           & 158.9 & 139 & 14.3 & 155.3 & 117 & 32.7 \\
Q4           & 160.0 & 139 & 15.1 & 156.5 & 117 & 33.8 \\
Q5 (highest) & 161.1 & 139 & 15.9 & 157.8 & 117 & 34.8 \\
\midrule
Q5$-$Q1 gap  & 4.0 & --- & 2.9 & 4.4 & --- & 3.8 \\
\bottomrule
\end{tabular}
\label{tab:fair_cm_2020}
\end{table}

Table~\ref{tab:fair_cm_2020} reports the main cross-sectional results for the reference year $t=2020$ at retirement ages $x=60$ and $x=63$. The pattern is clear. First, the official counting month is below the actuarially fair level for every income quintile, so the current rule generates positive subsidies across the board. Second, the subsidy is increasing in income, because higher-income groups have lower mortality and therefore larger annuity values. At age $60$, the fair counting month ranges from about $157$ months in Q1 to about $161$ months in Q5, compared with the official value $M_{60}^{\mathrm{off}}=139$. The corresponding subsidy rate rises from roughly $13\%$ to $15.9\%$. At age $63$, the official value is $M_{63}^{\mathrm{off}}=117$, whereas the fair counting month ranges from about $153.3$ to $157.7$ months, implying substantially larger subsidies of roughly $31.0\%$ to $34.8\%$. Thus, the current age-only conversion rule creates not only an overall actuarial shortfall, but also a reverse transfer from shorter-lived low-income retirees to longer-lived high-income retirees.

To examine whether this pattern is transitory or persistent, we project the common Lee--Carter time index $\kappa_t$ forward using a random walk with drift and recompute the fair counting month and subsidy rate for $t=2021,\ldots,2040$. The results reported below are medians over 1{,}000 simulation paths. Figure~\ref{fig:cm_subsidy_projection} shows that the problem is persistent. The actuarially fair counting month rises gradually over time for all quintiles, reflecting continuing mortality improvement, while the official counting month remains fixed at the statutory value for each retirement age. As a result, the subsidy rate also rises over time. The income gradient remains ordered throughout the projection horizon, with higher-income groups consistently receiving a larger subsidy than lower-income groups. In other words, absent reform of the conversion rule, mortality improvement tends to reinforce rather than eliminate the reverse transfer embedded in the current NDC design.

\begin{figure}[!ht]
    \centering
    \begin{subfigure}{0.49\textwidth}
        \centering
        \includegraphics[width=\linewidth]{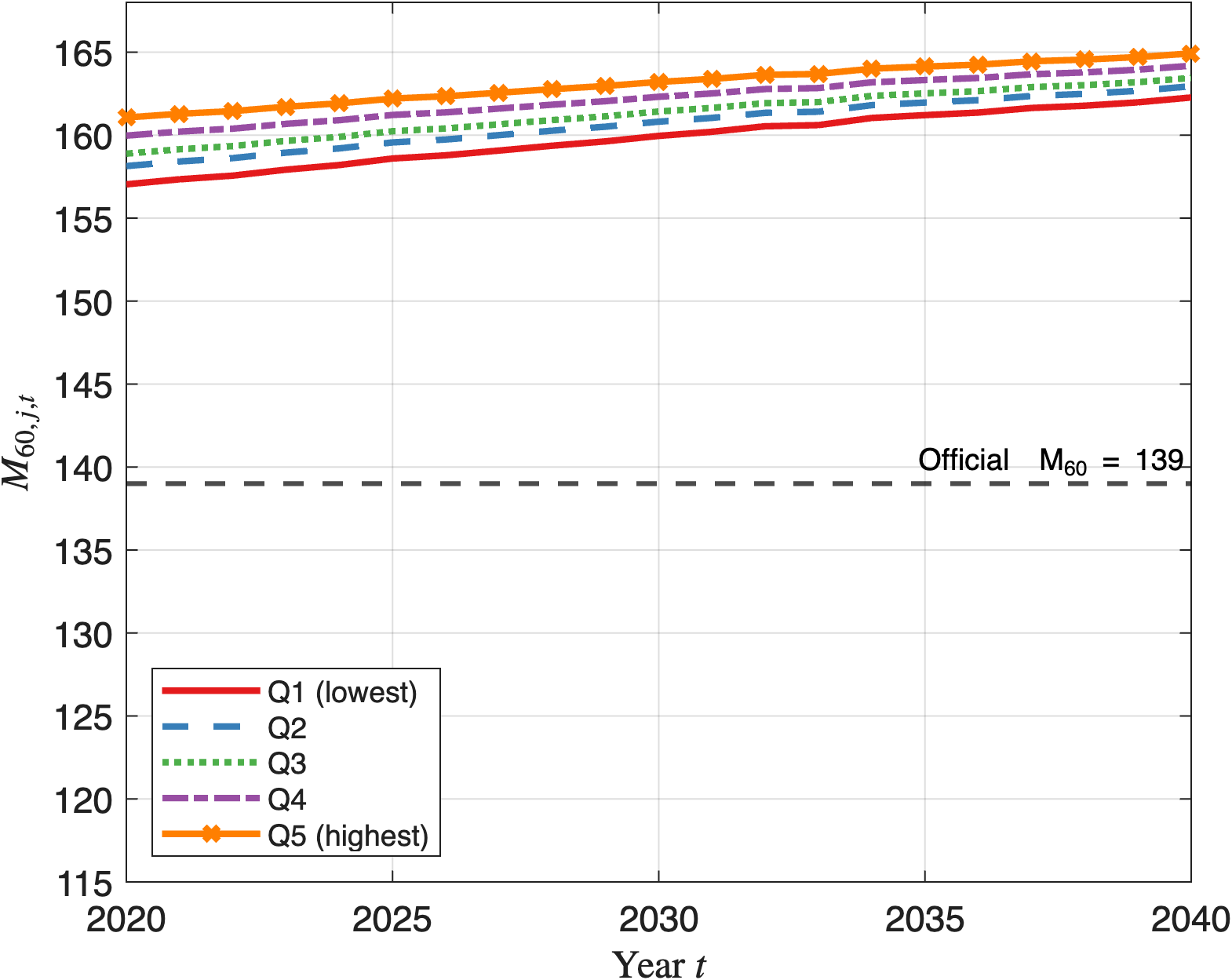}
        \caption{Age 60}
    \end{subfigure}
    \hfill
    \begin{subfigure}{0.49\textwidth}
        \centering
        \includegraphics[width=\linewidth]{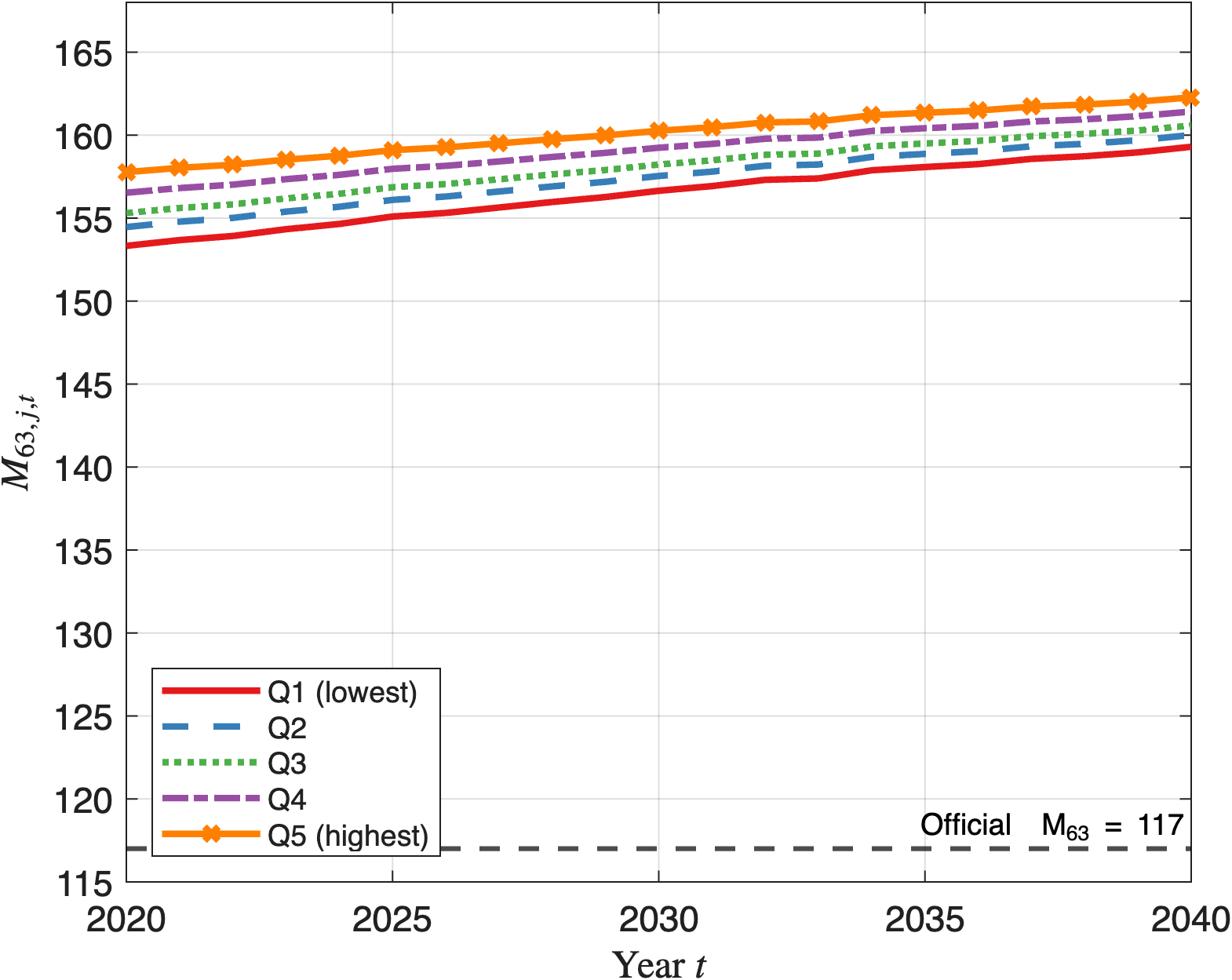}
        \caption{Age 63}
    \end{subfigure}
    
    \vspace{0.5em}

    \begin{subfigure}{0.49\textwidth}
        \centering
        \includegraphics[width=\linewidth]{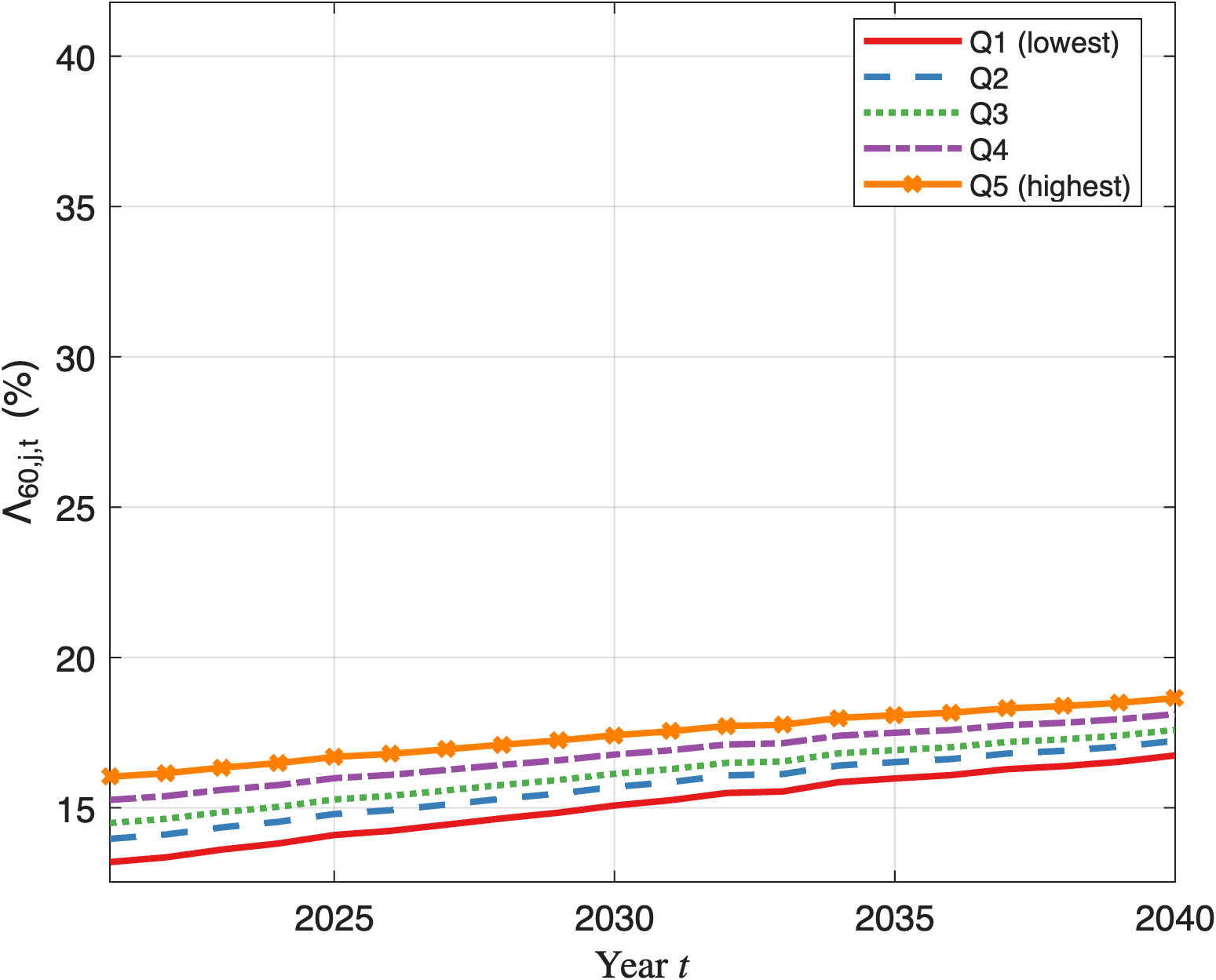}
        \caption{Age 60}
    \end{subfigure}
    \hfill
    \begin{subfigure}{0.49\textwidth}
        \centering
        \includegraphics[width=\linewidth]{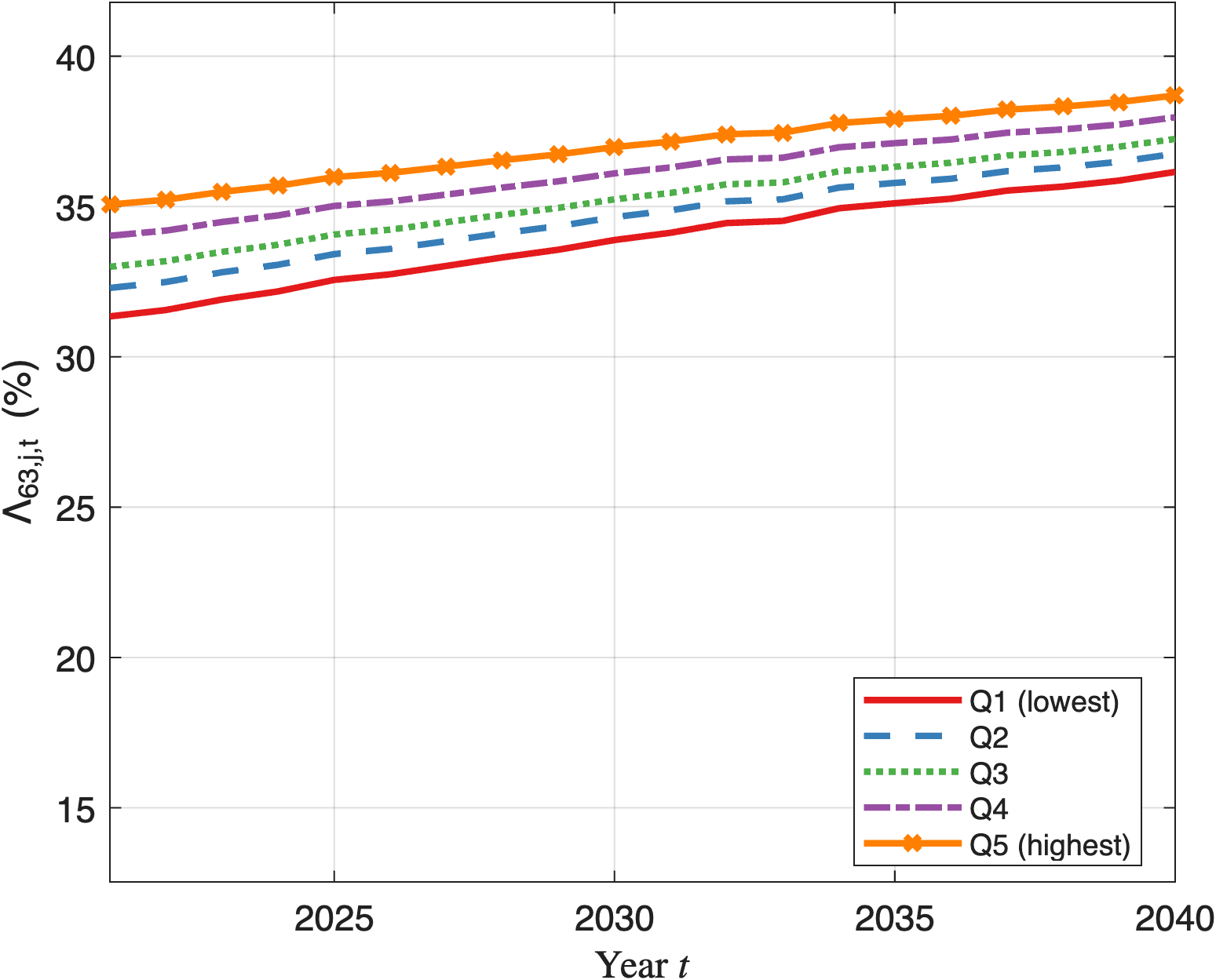}
        \caption{Age 63}
    \end{subfigure}
    \caption{Actuarially fair counting month $M^{\mathrm{fair}}_{x,j,t}$ (top row) and subsidy rate $\Lambda_{x,j,t}$ (bottom row) at years $t=2020-2040$ by income quintile.}
    \label{fig:cm_subsidy_projection}
\end{figure}

\subsection{Four progressive annuitization methods}
\label{subsec:four_methods}

We next consider income-dependent annuitization rules designed to reduce the reverse transfer documented above. Throughout this subsection, we focus on retirement at age 60 in year 2020. Let $K$ denote career-average annual income, expressed in 2020 CNY. For comparison across rules, we assume that the notional account balance equals $\phi K$, where
$\phi = 0.08 \times 30 = 2.4$,
corresponding to an $8\%$ contribution rate and $30$ years of participation, abstracting from wage growth and notional interest accumulation. This scaling affects monthly benefits but not the implied counting months.

Using the income groups defined in the CHARLS sample, set the quintile mean incomes to
\begin{equation*}
(K_1,K_2,K_3,K_4,K_5)
=
(2{,}181,\;6{,}131,\;12{,}902,\;23{,}897,\;51{,}599),
\end{equation*}
and the inter-quintile boundaries to
\begin{equation*}
(\bar K_0, \bar K_1,\bar K_2,\bar K_3,\bar K_4, \bar K_5)
=
(0, \; 3{,}847,\;8{,}838,\;17{,}651,\;31{,}300, \; \infty).
\end{equation*}
Let $M_j^\ast := M^{\mathrm{fair}}_{60,j,2020}$, $j=1,\ldots,5$, denote the actuarially fair counting month for quintile $j$ at age $60$ in year $2020$. In what follows, \(K_j\) and \(\bar K_j\) play different roles. The value \(K_j\) is the representative income of quintile \(j\), measured by the quintile mean, and is used as an actuarial anchor point. The value \(\bar K_j\) is the income cutoff between quintiles \(j\) and \(j+1\). We set \(\bar K_0=0\) and \(\bar K_5=\infty\), so bracket \(j\) is \((\bar K_{j-1},\bar K_j]\). By construction, \(K_j\in(\bar K_{j-1},\bar K_j]\) for \(j=1,\ldots,5\). Thus, in the bracket-based methods below, \(K_j-\bar K_{j-1}\) denotes the distance from the lower boundary of bracket \(j\) to its representative anchor income, while \(\bar K_j-\bar K_{j-1}\) denotes the full width of bracket \(j\). In the interpolation-based methods, the quintile means \(K_1,\ldots,K_5\) serve as the anchor grid.

\subsubsection{Continuous actuarially fair benchmark}

We first construct a continuous benchmark schedule over income. Specifically, for each age $x$, the estimated baseline log-mortality $\alpha_{x,j}$ is linearly interpolated across the quintile mean incomes and then clamped outside the observed range:
\begin{equation}
\alpha_x(K)
=
\begin{cases}
\alpha_{x,1}, & K \le K_1,\\[6pt]
\alpha_{x,j}
+ \dfrac{K-K_j}{K_{j+1}-K_j}\bigl(\alpha_{x,j+1}-\alpha_{x,j}\bigr),
& K_j < K \le K_{j+1},\quad j=1,\ldots,4,\\[10pt]
\alpha_{x,5}, & K > K_5.
\end{cases}
\label{eq:alpha_continuous}
\end{equation}
Using this interpolated mortality surface, we define the continuously actuarially fair counting month
\begin{equation*}
M^{\mathrm{fair}}(K) := 12\,\ddot{a}_{60,2020}^{(12)}(K).
\end{equation*}
This object is the actuarially fair benchmark against which the four implementable methods are compared.

For any proposed rule, let $M(K)$ denote the implied average counting month. The corresponding monthly pension benefit is
$B(K)=\frac{\phi K}{M(K)}$.
When a rule is specified in marginal terms, the associated marginal counting month is
\begin{equation*}
\delta(K)
=
\left[
\frac{d}{dK}\left(\frac{K}{M(K)}\right)
\right]^{-1}
=
\left[
\frac{d}{dK}\left(\frac{B(K)}{\phi}\right)
\right]^{-1}.
\end{equation*}
Note that $\delta(K)$ has the same unit as the counting month.\footnote{
Denote by $\delta(K,\Delta)$ the marginal counting month applied to the additional income between $K$ and $K+\Delta$, where the additional account balance is $\phi\Delta$. It is natural to define the marginal counting month as the limit as $\Delta\rightarrow 0^+$. By the definition of the monthly benefit,
\begin{align*}
B(K+\Delta)
=
B(K)+\frac{\phi\Delta}{\delta(K,\Delta)}
\quad \Longrightarrow \quad
\delta(K)
&=
\lim_{\Delta\rightarrow 0^+}\delta(K,\Delta)  \\
&=
\lim_{\Delta\rightarrow 0^+}
\left(
\frac{B(K+\Delta)-B(K)}{\phi\Delta}
\right)^{-1}
=
\left[
\frac{d}{dK}\left(\frac{B(K)}{\phi}\right)
\right]^{-1}.
\end{align*}
} 
Its reciprocal is the marginal conversion rate.

\subsubsection{Method 1: piecewise-constant average counting month}

Method 1 is the most direct extension of the current rule. It assigns a constant average counting month to each income bracket:
\begin{equation}
M_1(K)
=
\begin{cases}
M_1^\ast, & 0 \le K \le \bar K_1,\\
M_2^\ast, & \bar K_1 < K \le \bar K_2,\\
M_3^\ast, & \bar K_2 < K \le \bar K_3,\\
M_4^\ast, & \bar K_3 < K \le \bar K_4,\\
M_5^\ast, & K > \bar K_4.
\end{cases}
\label{eq:method1}
\end{equation}
The corresponding monthly benefit is $B_1(K)=\phi K/M_1(K)$. This rule is simple, but it inherits the main drawback of a step schedule: at an income threshold, the divisor jumps discretely, so total monthly pension need not be continuous and may even fall locally when $K$ crosses a boundary.

\subsubsection{Method 2: piecewise-linear average counting month}

Method 2 smooths the previous rule by linearly interpolating the average counting month across the quintile mean incomes:
\begin{equation}
M_2(K)
=
\begin{cases}
M_1^\ast, & K \le K_1,\\[6pt]
M_j^\ast + \dfrac{M_{j+1}^\ast-M_j^\ast}{K_{j+1}-K_j}(K-K_j),
& K_j < K \le K_{j+1},\quad j=1,\ldots,4,\\[10pt]
M_5^\ast, & K > K_5.
\end{cases}
\label{eq:method2}
\end{equation}
The corresponding monthly benefit is $B_2(K)=\phi K/M_2(K)$. This method preserves continuity of both the divisor and the total monthly pension. However, it does not directly control the marginal treatment of additional income: its implied marginal counting month is
\[
\delta_2(K)
=
\left[\frac{d}{dK}\left(\frac{K}{M_2(K)}\right)\right]^{-1}
=
\frac{M_2(K)^2}{M_2(K)-K M_2'(K)},
\]
which is a complicated and nonlinear function of $K$. Thus, a linear rule for the average counting month generally induces a different marginal schedule.

A simple example using the last interpolation segment illustrates the issue. On the interval $[K_4,K_5]=[23{,}897,\;51{,}599]$,
Method 2 linearly interpolates the actuarially fair anchor values $M_4^\ast=160.0$ and $M_5^\ast=161.1$.
Consider two individuals whose career-average incomes differ by 100 CNY, one at $K=51{,}599$ and the other at $K=51{,}499$. Method 2 assigns
\[
M_2(51{,}599)=161.1,
\qquad
M_2(51{,}499)\approx 161.0960.
\]
The counting month applied to the additional 100 CNY of career-average income is therefore
\[
\frac{100}{
51{,}599/M_2(51{,}599)-51{,}499/M_2(51{,}499)
}
\approx 163.17.
\]
By contrast, for $K>K_5$, Method 2 fixes the average divisor at 161.1, so the next additional 100 CNY above $K_5$ is converted at marginal counting month 161.1. Hence the implied marginal counting month falls discretely from about 163.17 just below $K_5$ to 161.1 just above $K_5$. Total monthly benefits remain continuous and increasing, but the marginal treatment of additional income becomes more favorable after crossing the knot. This feature motivates Methods 3 and 4, which specify the reform directly in terms of marginal counting months rather than indirectly through the average divisor.

\subsubsection{Method 3: piecewise-constant marginal counting month (bracket formula)}

Method 3 keeps a bracket structure, but applies it to the \emph{marginal} treatment of additional earnings rather than to the average divisor itself. The idea is analogous to a bracket-based tax schedule. Each income bracket has its own marginal counting month, and total monthly benefits are obtained by accumulating the converted benefits from all brackets below the individual’s income level. Thus, the rule governs how each additional unit of career-average income is converted into pension benefits, rather than assigning one average divisor to the entire account balance.

A natural first approach would be to choose the bracket-specific marginal counting months so that the benefit schedule passes exactly through the actuarially fair anchor points. Let
\[
q_j:=\frac{K_j}{M_j^\ast},
\qquad j=1,\ldots,5,
\]
denote the actuarially fair normalized benefit at anchor income \(K_j\). The normalization by \(\phi\) is useful because the account balance is \(\phi K\), so a monthly benefit \(B(K)\) corresponds to the normalized benefit \(B(K)/\phi\).

Suppose exact anchor matching were imposed. At the beginning of bracket \(j\), the normalized benefit already accumulated from lower brackets is \(C_{j-1}^{\mathrm{exact}}\). Within bracket \(j\), each additional unit of income is converted at marginal counting month \(\delta_{3,j}^{\mathrm{exact}}\), so the additional normalized benefit from moving from \(\bar K_{j-1}\) to \(K_j\) is
\[
\frac{K_j-\bar K_{j-1}}{\delta_{3,j}^{\mathrm{exact}}}.
\]
Exact matching at \(K_j\) therefore requires
\[
C_{j-1}^{\mathrm{exact}}
+
\frac{K_j-\bar K_{j-1}}{\delta_{3,j}^{\mathrm{exact}}}
=
q_j.
\]
Solving this equation gives
\[
\delta_{3,j}^{\mathrm{exact}}
=
\frac{K_j-\bar K_{j-1}}{q_j-C_{j-1}^{\mathrm{exact}}},
\qquad
j=1,\ldots,5.
\]
The cumulative benefit at the upper boundary of bracket \(j\) is then updated as
\[
C_j^{\mathrm{exact}}
=
C_{j-1}^{\mathrm{exact}}
+
\frac{\bar K_j-\bar K_{j-1}}{\delta_{3,j}^{\mathrm{exact}}},
\qquad
j=1,\ldots,4,
\]
with \(C_0^{\mathrm{exact}}=0\). Thus, \(C_j^{\mathrm{exact}}\) is simply the normalized benefit accumulated up to the end of bracket \(j\).

This exact-matching formula is useful for interpretation, but it need not produce a monotone marginal schedule. For \(j\ge2\), exact matching is consistent with a weakly increasing marginal counting month only if
\[
C_{j-1}^{\mathrm{exact}}
<
q_j
\le
C_{j-1}^{\mathrm{exact}}
+
\frac{K_j-\bar K_{j-1}}{\delta_{3,j-1}^{\mathrm{exact}}}.
\]
The first inequality ensures that the remaining benefit \(q_j-C_{j-1}^{\mathrm{exact}}\) is positive, so that a positive marginal counting month can be defined. The second inequality ensures that the new marginal counting month is not smaller than the previous one. To see this, note that
\[
\delta_{3,j}^{\mathrm{exact}}\ge \delta_{3,j-1}^{\mathrm{exact}}
\quad \Longleftrightarrow \quad
\frac{K_j-\bar K_{j-1}}{q_j-C_{j-1}^{\mathrm{exact}}}
\ge
\delta_{3,j-1}^{\mathrm{exact}},
\]
which is equivalent to the upper bound above. If \(q_j\) is too large relative to the benefit already accumulated from previous brackets, then exact matching requires a smaller \(\delta_{3,j}^{\mathrm{exact}}\), meaning a larger marginal conversion rate at a higher income level. Conversely, if \(q_j\le C_{j-1}^{\mathrm{exact}}\), the previous brackets have already generated at least as much benefit as the fair anchor value, so exact matching with a positive marginal counting month is infeasible.

For this reason, we do not impose exact fairness at all anchor points for Method 3. Instead, we choose the monotone marginal schedule that comes closest to the fair anchor values. For any candidate vector
\[
\delta_3=(\delta_{3,1},\ldots,\delta_{3,5})
\]
satisfying
\[
0<\delta_{3,1}\le\delta_{3,2}\le\cdots\le\delta_{3,5},
\]
define
\[
C_0(\delta_3)=0,
\qquad
C_j(\delta_3)
=
C_{j-1}(\delta_3)
+
\frac{\bar K_j-\bar K_{j-1}}{\delta_{3,j}},
\qquad
j=1,\ldots,4.
\]
For a given \(\delta_3\), the normalized benefit delivered at anchor \(K_j\) is
\[
Y_{3,j}(\delta_3)
=
C_{j-1}(\delta_3)
+
\frac{K_j-\bar K_{j-1}}{\delta_{3,j}},
\qquad
j=1,\ldots,5.
\]
This expression has the same accumulation logic as above: the first term is the benefit accumulated from previous brackets, and the second term is the benefit generated within bracket \(j\) up to the anchor \(K_j\).

We then determine the marginal counting months by solving
\begin{equation}
\widehat\delta_3
=
\underset{{0<\delta_{3,1}\le\delta_{3,2}\le\cdots\le\delta_{3,5}}}{\arg\min}
\sum_{j=1}^5
\left(
\frac{Y_{3,j}(\delta_3)}{q_j}-1
\right)^2 .
\label{eq:method3_optimization}
\end{equation}
The objective function measures the squared proportional deviations from the actuarially fair normalized benefits at the anchor incomes. Thus, Method 3 chooses the marginal counting months that come as close as possible to actuarial fairness at the anchors, while enforcing weakly increasing marginal counting months over income. This is a constrained nonlinear least-squares problem.

Given the calibrated vector
\[
\widehat{\delta}_3=(\widehat{\delta}_{3,1},\ldots,\widehat{\delta}_{3,5}),
\]
define
\[
\widehat C_0=0,
\qquad
\widehat C_j
=
\widehat C_{j-1}
+
\frac{\bar K_j-\bar K_{j-1}}{\widehat\delta_{3,j}},
\qquad
j=1,\ldots,4.
\]
The resulting benefit schedule is
\begin{equation}
\frac{B_3(K)}{\phi}
=
\begin{cases}
\dfrac{K}{\widehat\delta_{3,1}}, & 0 \le K \le \bar K_1,\\[8pt]
\widehat C_{j-1}
+
\dfrac{K-\bar K_{j-1}}{\widehat\delta_{3,j}},
& \bar K_{j-1}<K\le \bar K_j,\quad j=2,3,4,\\[10pt]
\widehat C_4+\dfrac{K-\bar K_4}{\widehat\delta_{3,5}}, & K>\bar K_4.
\end{cases}
\label{eq:method3}
\end{equation}
The implied average counting month is
\[
M_3(K)=\frac{\phi K}{B_3(K)}.
\]

The need for the monotonicity-constrained calibration is visible in the current data. If we impose exact anchor matching using the age-60 anchor values, the resulting bracket-specific marginal counting months are approximately
\[
(\delta_{3,1}^{\mathrm{exact}},\ldots,\delta_{3,5}^{\mathrm{exact}})
=
(157.0,\;160.0,\;159.4,\;162.8,\;161.8).
\]
Although this exact-matching schedule reproduces the five fair anchor values, it is not globally monotone: the marginal counting month falls from the second to the third bracket and again from the fourth to the fifth bracket. In those ranges, the marginal conversion rate would increase with income, which is inconsistent with the intended progressive marginal interpretation. The constrained calibration in \eqref{eq:method3_optimization} therefore replaces the exact-matching schedule with the closest weakly increasing sequence of marginal counting months.

\subsubsection{Method 4: piecewise-linear marginal counting month}

Method 4 smooths Method 3 by defining the marginal counting month on the anchor-income grid \(K_1,\ldots,K_5\), rather than on the bracket-boundary grid \(\bar K_0,\ldots,\bar K_5\). The knot value \(\delta_{4,j}\) is the marginal counting month at the representative income \(K_j\), and \(\delta_4(K)\) is linearly interpolated between adjacent anchor incomes. The motivation is to keep the marginal-rule interpretation of Method 3 while avoiding sharp jumps in the marginal counting month at bracket boundaries. Thus, Method 4 produces a continuous marginal counting-month schedule: nearby income levels face nearby marginal conversion rates.

As in Method 3, let
\[
q_j:=\frac{K_j}{M_j^\ast},
\qquad j=1,\ldots,5,
\]
denote the actuarially fair normalized benefit at anchor income \(K_j\). Recall that the marginal counting month is defined by
\[
\delta(K)
=
\left[
\frac{d}{dK}\left(\frac{B(K)}{\phi}\right)
\right]^{-1}.
\]
Equivalently,
\[
\frac{d}{dK}\left(\frac{B(K)}{\phi}\right)
=
\frac{1}{\delta(K)}.
\]
Thus, \(1/\delta(K)\) is the marginal conversion rate: it tells us how much normalized monthly benefit is generated by one additional unit of career-average income. Consequently, once a marginal counting-month schedule \(\delta_4(K)\) is specified, the normalized benefit \(B_4(K)/\phi\) is obtained by integrating the marginal conversion rate \(1/\delta_4(K)\) over income.

A natural first approach would again be exact anchor matching. At the first anchor, exact matching would require
\[
\frac{B_4(K_1)}{\phi}=q_1.
\]
Since Method 4 uses a constant marginal counting month below \(K_1\), this implies
\[
\frac{K_1}{\delta_{4,1}^{\mathrm{exact}}}=q_1
\quad\Longrightarrow\quad
\delta_{4,1}^{\mathrm{exact}}=M_1^\ast.
\]
For subsequent anchors, suppose the rule has already matched \(q_j\) at \(K_j\), and the current marginal counting month at that point is \(\delta_{4,j}^{\mathrm{exact}}\). To match the next anchor \(q_{j+1}\), the normalized benefit generated between \(K_j\) and \(K_{j+1}\) must equal \(q_{j+1}-q_j\), that is,
\[
q_{j+1}-q_j
=
\int_{K_j}^{K_{j+1}}\frac{1}{\delta_4(C)}\,dC.
\]
This equation determines the next knot value \(\delta_{4,j+1}^{\mathrm{exact}}\). However, exact anchor matching can conflict with monotonicity. If \(\delta_4(K)\) is required to be weakly increasing on \((K_j,K_{j+1}]\), then \(\delta_4(C)\ge \delta_{4,j}\) over this interval, and therefore
\[
\frac{1}{\delta_4(C)}
\le
\frac{1}{\delta_{4,j}}.
\]
Integrating both sides gives
\[
\int_{K_j}^{K_{j+1}}\frac{1}{\delta_4(C)}\,dC
\le
\frac{K_{j+1}-K_j}{\delta_{4,j}}.
\]
Therefore, exact matching of the next anchor is compatible with a nondecreasing marginal counting month only if
\[
0
<
q_{j+1}-q_j
\le
\frac{K_{j+1}-K_j}{\delta_{4,j}}.
\]
If the target increment \(q_{j+1}-q_j\) is too large, then the average marginal conversion rate required between \(K_j\) and \(K_{j+1}\) is too high to be achieved by a nondecreasing marginal counting month starting from \(\delta_{4,j}\). In that case, exact matching would require \(\delta_4(K)\) to fall over the interval, which is inconsistent with the intended progressive marginal interpretation.

For this reason, Method 4 treats the fair anchor values as targets rather than hard constraints. Let
\[
\delta_4=(\delta_{4,1},\ldots,\delta_{4,5})
\]
denote a candidate vector of marginal counting-month knots. We impose
\[
0<\delta_{4,1}\le\delta_{4,2}\le\cdots\le\delta_{4,5},
\]
which ensures that the marginal counting month is weakly increasing over income. Since \(\delta_4(K)\) is linearly interpolated between adjacent knots, this constraint is equivalent to \(\delta_4'(K)\ge0\) on each interpolation interval.

For any candidate knot vector \(\delta_4\), define the piecewise-linear marginal counting month by
\[
\delta_4(K)
=
\frac{K_{j+1}-K}{K_{j+1}-K_j}\,\delta_{4,j}
+
\frac{K-K_j}{K_{j+1}-K_j}\,\delta_{4,j+1},
\qquad
K_j<K\le K_{j+1},\quad j=1,\ldots,4.
\]
This expression is simply linear interpolation. The first weight is one at \(K_j\) and zero at \(K_{j+1}\), while the second weight is zero at \(K_j\) and one at \(K_{j+1}\). Hence \(\delta_4(K_j)=\delta_{4,j}\) and \(\delta_4(K_{j+1})=\delta_{4,j+1}\).

Next define
\[
I_j(\delta_{4,j},\delta_{4,j+1})
:=
\int_{K_j}^{K_{j+1}}\frac{1}{\delta_4(C)}\,dC.
\]
The quantity \(I_j(\delta_{4,j},\delta_{4,j+1})\) is the total normalized benefit generated by income between \(K_j\) and \(K_{j+1}\). It is the integral of the marginal conversion rate over that income interval. Because \(\delta_4(K)\) is linear on each interval, this integral has the closed form
\[
I_j(a,b)
=
\begin{cases}
\dfrac{K_{j+1}-K_j}{a}, & a=b,\\[10pt]
\dfrac{K_{j+1}-K_j}{b-a}\log\!\left(\dfrac{b}{a}\right), & a\ne b.
\end{cases}
\]
The first case applies when the marginal counting month is constant over the interval. The second case applies when the marginal counting month changes linearly from \(a\) to \(b\).

For any candidate \(\delta_4\), define the normalized benefit at the anchor incomes recursively. For the first anchor,
\[
Y_{4,1}(\delta_4)=\frac{K_1}{\delta_{4,1}}.
\]
This is the normalized benefit accumulated from zero income to \(K_1\), using the marginal counting month \(\delta_{4,1}\). For subsequent anchors,
\[
Y_{4,j+1}(\delta_4)
=
Y_{4,j}(\delta_4)
+
I_j(\delta_{4,j},\delta_{4,j+1}),
\qquad
j=1,\ldots,4.
\]
Thus, \(Y_{4,j}(\delta_4)\) is the normalized benefit \(B_4(K_j)/\phi\) implied by the candidate marginal schedule at anchor income \(K_j\). The recursion says that the benefit at \(K_{j+1}\) equals the benefit already accumulated at \(K_j\) plus the incremental benefit generated over the interval \([K_j,K_{j+1}]\).

The knot values are chosen by solving
\begin{equation}
\widehat\delta_4
=
\argmin_{0<\delta_{4,1}\le\delta_{4,2}\le\cdots\le\delta_{4,5}}
\sum_{j=1}^5
\left(
\frac{Y_{4,j}(\delta_4)}{q_j}-1
\right)^2 .
\label{eq:method4_optimization}
\end{equation}
The objective compares the normalized benefit generated by the candidate rule, \(Y_{4,j}(\delta_4)\), with the actuarially fair normalized benefit \(q_j\) at each anchor. Therefore, \((Y_{4,j}(\delta_4)/q_j-1)\) is the proportional deviation from actuarial fairness at anchor \(K_j\). Method 4 chooses the monotone piecewise-linear marginal counting-month schedule that minimizes the squared proportional deviations across the five anchors.

Given the calibrated knot vector
$\widehat\delta_4=(\widehat\delta_{4,1},\ldots,\widehat\delta_{4,5})$,
define the fitted anchor benefits by
\[
\widehat Y_{4,1}:=\frac{K_1}{\widehat\delta_{4,1}},
\qquad
\widehat Y_{4,j+1}
=
\widehat Y_{4,j}
+
I_j(\widehat\delta_{4,j},\widehat\delta_{4,j+1}),
\quad j=1,\ldots,4.
\]
For \(K_j<K\le K_{j+1}\), the fitted marginal counting month is
\[
\widehat\delta_4(K)
=
\frac{K_{j+1}-K}{K_{j+1}-K_j}\,\widehat\delta_{4,j}
+
\frac{K-K_j}{K_{j+1}-K_j}\,\widehat\delta_{4,j+1}.
\]
The resulting benefit schedule is
\begin{equation}
\frac{B_4(K)}{\phi}
=
\begin{cases}
\dfrac{K}{\widehat\delta_{4,1}}, & K \le K_1,\\[8pt]
\widehat Y_{4,j}+\displaystyle\int_{K_j}^{K}\frac{1}{\widehat\delta_4(C)}\,dC,
& K_j < K \le K_{j+1},\quad j=1,\ldots,4,\\[12pt]
\widehat Y_{4,5}+\dfrac{K-K_5}{\widehat\delta_{4,5}}, & K > K_5.
\end{cases}
\label{eq:method4}
\end{equation}
This formula again follows from accumulation. Below \(K_1\), income is converted using the first marginal counting month. Between \(K_j\) and \(K_{j+1}\), the rule starts from the benefit already accumulated at \(K_j\), namely \(\widehat Y_{4,j}\), and adds the marginal benefit generated between \(K_j\) and \(K\). Above \(K_5\), the marginal counting month is held fixed at its final value \(\widehat\delta_{4,5}\).

The implied average counting month is
\begin{equation*}
M_4(K)=\frac{\phi K}{B_4(K)}.
\end{equation*}
Thus, Method 4 specifies the marginal counting month directly, and the average counting month is an outcome of the accumulated marginal conversion rates.

For implementation, the partial integral in \eqref{eq:method4} is
\[
\int_{K_j}^{K}\frac{1}{\widehat\delta_4(C)}\,dC
=
\begin{cases}
\dfrac{K-K_j}{\widehat\delta_{4,j}}, 
& \widehat\delta_{4,j+1}=\widehat\delta_{4,j},\\[10pt]
\dfrac{K_{j+1}-K_j}{\widehat\delta_{4,j+1}-\widehat\delta_{4,j}}
\log\!\left[
\frac{\widehat\delta_{4,j}
+
(\widehat\delta_{4,j+1}-\widehat\delta_{4,j})
\dfrac{K-K_j}{K_{j+1}-K_j}}
{\widehat\delta_{4,j}}
\right],
& \widehat\delta_{4,j+1}\ne \widehat\delta_{4,j}.
\end{cases}
\]
This expression is used to evaluate the benefit schedule at income levels between two adjacent anchors.

As in Method 3, the constrained rule need not pass exactly through every actuarially fair anchor point. It instead chooses the smooth monotone marginal schedule that best approximates those anchors. If exact anchor matching is feasible under \(\delta_4'(K)\ge0\), then the objective in Optimization \eqref{eq:method4_optimization} is zero.

The need for the monotonicity-constrained calibration is again visible in the current data. If exact anchor matching is imposed using the age-60 values, the recursive knot values are approximately
\[
(\delta_{4,1}^{\mathrm{exact}},\ldots,\delta_{4,5}^{\mathrm{exact}})
=
(157.0,\;160.4,\;158.8,\;163.8,\;160.3).
\]
Similar to the example in Method 3, although this exact-matching schedule reproduces the five fair anchor values, it is not globally monotone: the marginal counting month falls from the second to the third knot and again from the fourth to the fifth knot. In those ranges, the marginal conversion rate would increase with income, which is inconsistent with the intended progressive marginal interpretation. The constrained calibration in \eqref{eq:method4_optimization} therefore replaces the exact-matching schedule with the closest weakly increasing piecewise-linear marginal counting-month schedule.

\subsubsection{Evaluation of the four methods}

For each method $m=1,\ldots,4$, define the residual subsidy at income level $K$ by
\begin{equation*}
\Lambda_m(K)
=
\frac{M^{\mathrm{fair}}(K)}{M_m(K)}-1.
\end{equation*}
Methods 1 and 2 match the actuarially fair benchmark exactly at the five anchor incomes:
\[
M_m(K_j)=M_j^\ast,
\qquad
\Lambda_m(K_j)=0,
\qquad
j=1,\ldots,5,\quad m=1,2.
\]
Methods 3 and 4 instead treat these anchor values as targets. They choose monotone marginal counting-month schedules that minimize proportional deviations from the fair normalized benefits \(q_j=K_j/M_j^\ast\). Therefore, their residual subsidies at the anchor points need not be exactly zero, but they remain small when the monotonicity constraint is not too restrictive.

Figure~\ref{fig:four_methods} compares the four rules in terms of monthly pension benefit, the benefit-to-income ratio, the implied average counting month, and the marginal counting month. Panel (a) shows that all four methods generate very similar monthly pension levels. This is because the actuarially fair counting months derived from the four proposed methods do not differ substantially across the income range considered here, so the four conversion rules imply similar total benefits at the scale of the plot. Panel (b) makes the differences more visible by reporting monthly benefit per 1,000 CNY of career-average income. The benefit-to-income ratio declines with income, reflecting the fairness adjustment: higher-income retirees have lower mortality and therefore receive a larger actuarially fair counting month, so each unit of account balance is converted into a slightly smaller monthly benefit.

\begin{figure}[!ht]
    \centering
    \begin{subfigure}{0.49\textwidth}
        \centering
        \includegraphics[width=\linewidth]{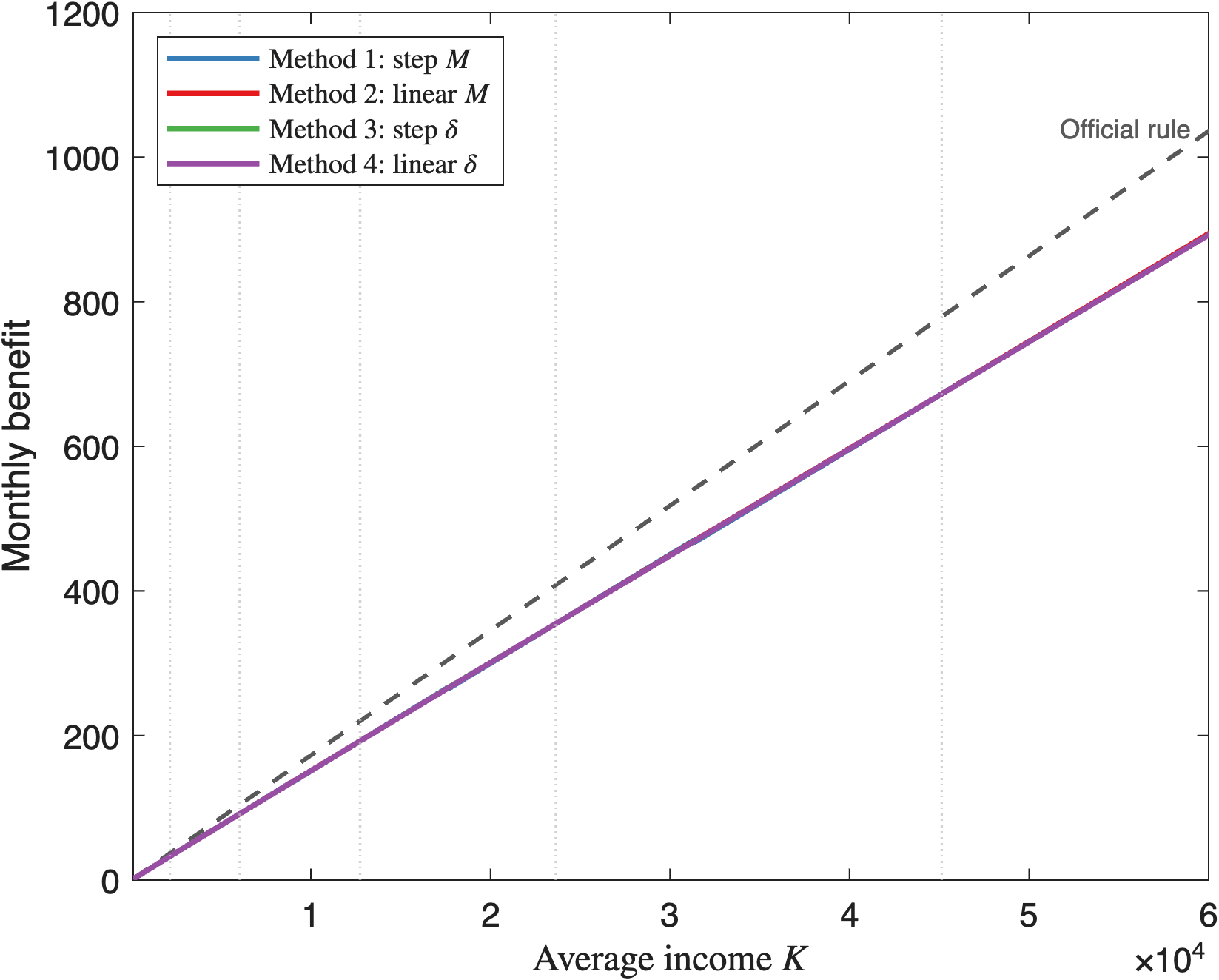}
        \caption{Monthly pension benefit}
    \end{subfigure}
    \hfill
    \begin{subfigure}{0.49\textwidth}
        \centering
        \includegraphics[width=\linewidth]{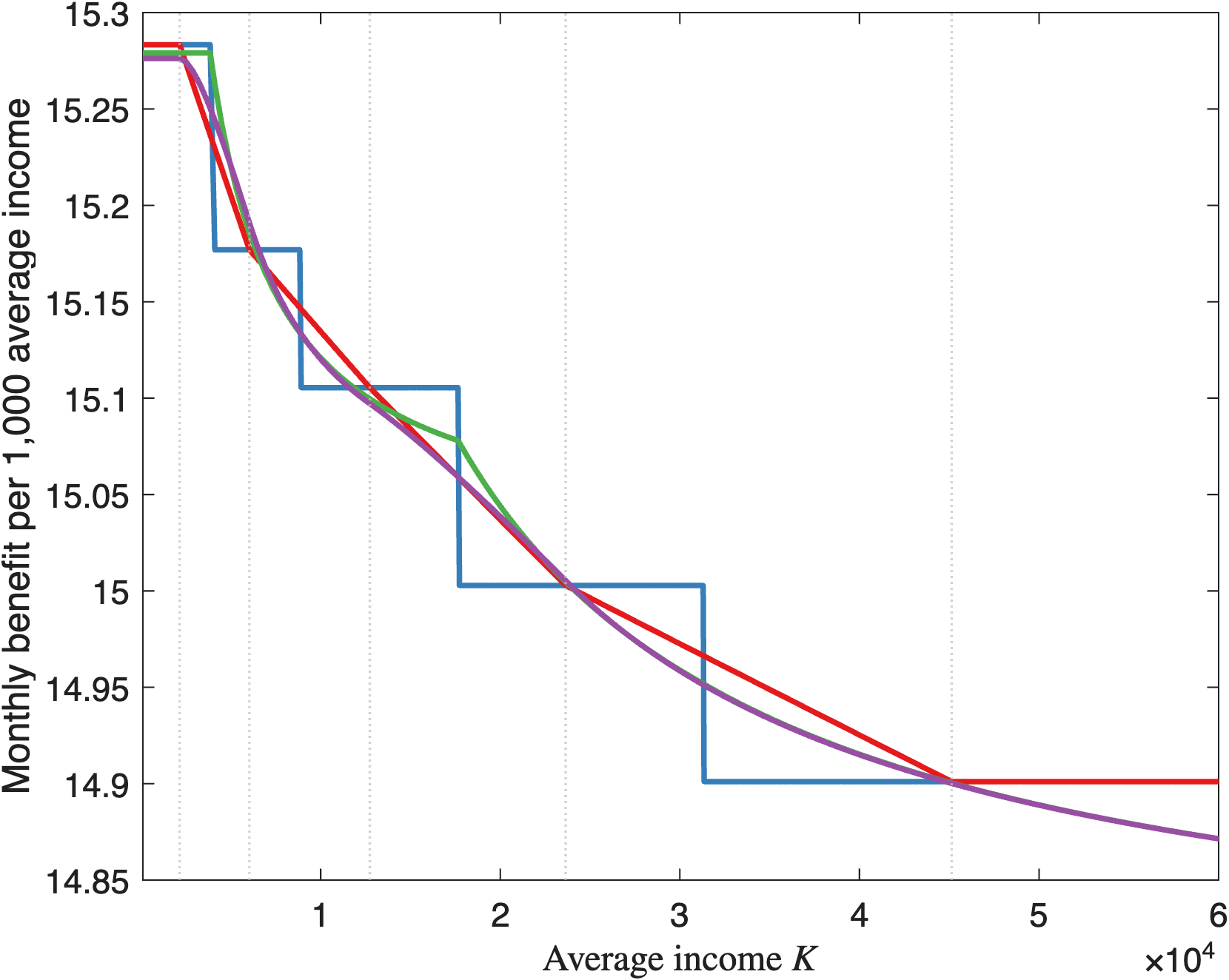}
        \caption{Benefit-to-income ratio}
    \end{subfigure}

    \vspace{0.5em}

    \begin{subfigure}{0.49\textwidth}
        \centering
        \includegraphics[width=\linewidth]{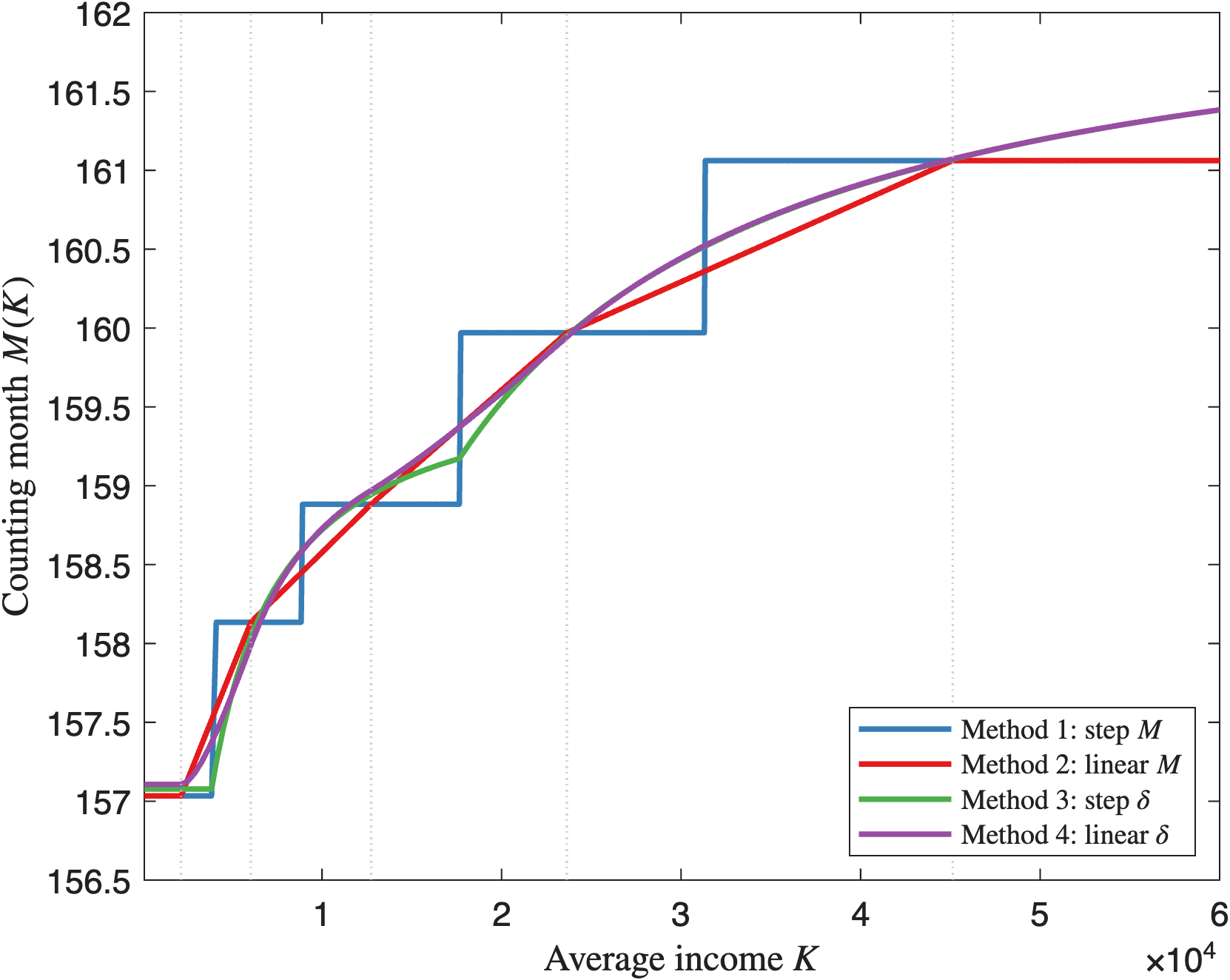}
        \caption{Implied counting month}
    \end{subfigure}
    \hfill
    \begin{subfigure}{0.49\textwidth}
        \centering
        \includegraphics[width=\linewidth]{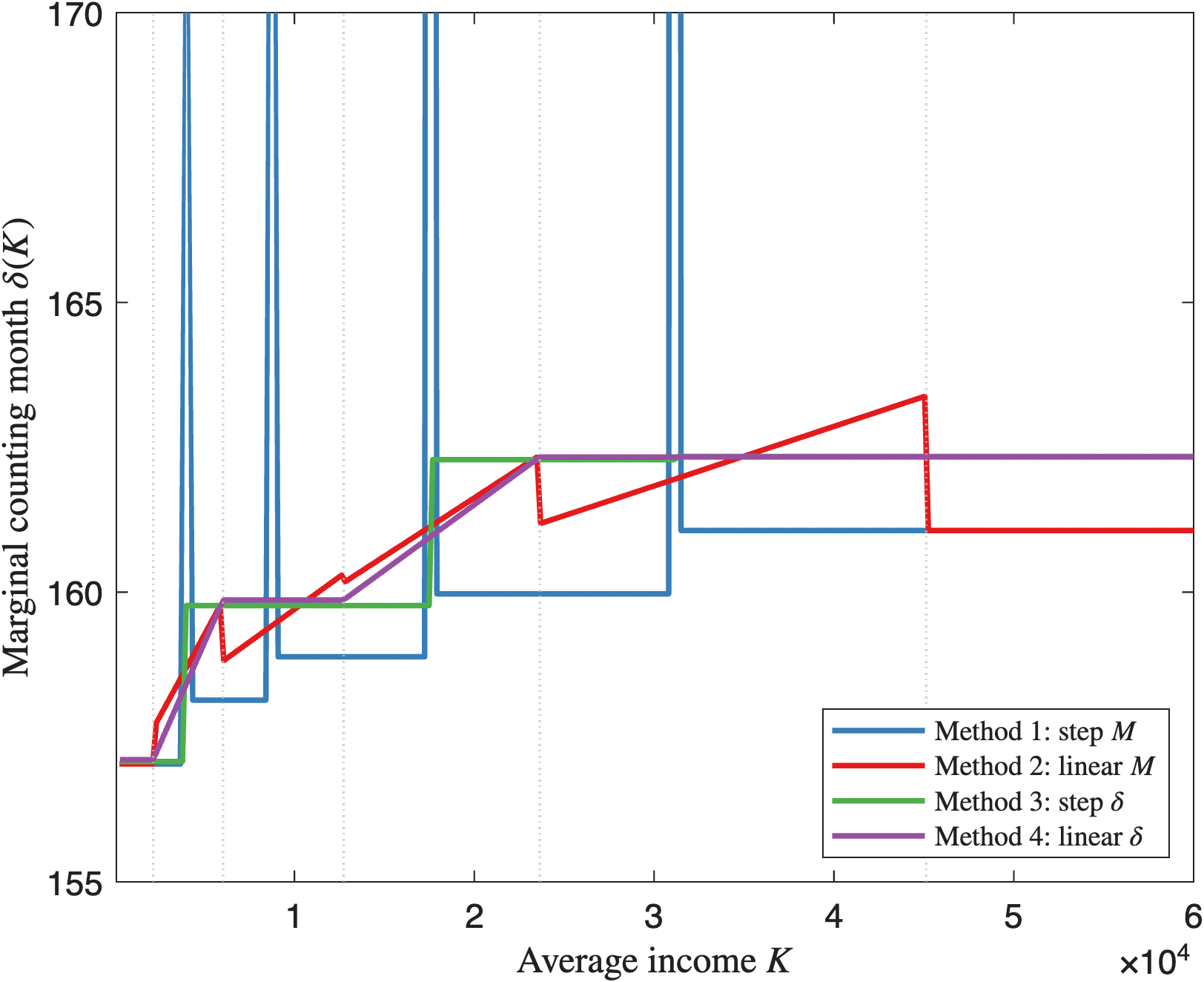}
        \caption{Implied marginal counting month}
    \end{subfigure}

    \caption{Comparison of the four pension conversion methods across career-average income \(K\) (2020 CNY).}
    \label{fig:four_methods}
\end{figure}

Panel (c) shows how the four methods implement this adjustment through the implied average counting month. Method 1 applies a bracket-based step rule to the average divisor. This makes the rule simple, but it creates discrete jumps at the income thresholds. Method 2 removes these jumps by linearly interpolating the average counting month across the quintile mean incomes. It therefore produces a smooth average-divisor schedule that closely tracks the continuously fair benchmark. Methods 3 and 4 behave differently because they are calibrated through marginal counting months rather than average counting months. Their implied average counting months are therefore outcomes of the accumulated marginal conversion rates. Since the monotonicity-constrained versions of Methods 3 and 4 treat the fair anchor values as targets rather than hard constraints, their implied average counting months do not need to pass exactly through every anchor point.

Panel (d) highlights the central distinction between average-divisor and marginal-divisor rules. For Method 1, the sharp movements in the marginal counting month reflect threshold notches: because the average divisor jumps at bracket boundaries, the marginal treatment of income around those boundaries is not well behaved. Method 2 smooths the average counting month, but its implied marginal counting month is still not directly controlled. Instead, it is an induced object and need not satisfy a simple monotonicity condition. By contrast, Methods 3 and 4 impose monotonicity directly on the marginal counting month. Method 3 produces a weakly increasing step schedule, analogous to a bracket-based marginal tax schedule. Method 4 further smooths this design by using a weakly increasing piecewise-linear marginal counting month. Thus, Methods 3 and 4 make explicit how each additional unit of career income is annuitized, while Method 4 avoids the sharp marginal jumps present in the step-based methods.

Figure~\ref{fig:residual_subsidy} shows the corresponding residual subsidy \(\Lambda_m(K)\). The first point to note is the scale of the remaining distortion. Under the current official age-only rule, the subsidy rate at age 60 ranges from about 13.0\% to 15.9\% across income quintiles. By contrast, after applying any of the four income-dependent rules, the residual subsidy remains close to zero over the income range shown in the figure. Thus, all four methods reduce the actuarial unfairness by more than an order of magnitude relative to the current official rule.

\begin{figure}[!ht]
    \centering
    \includegraphics[width=0.8\textwidth]{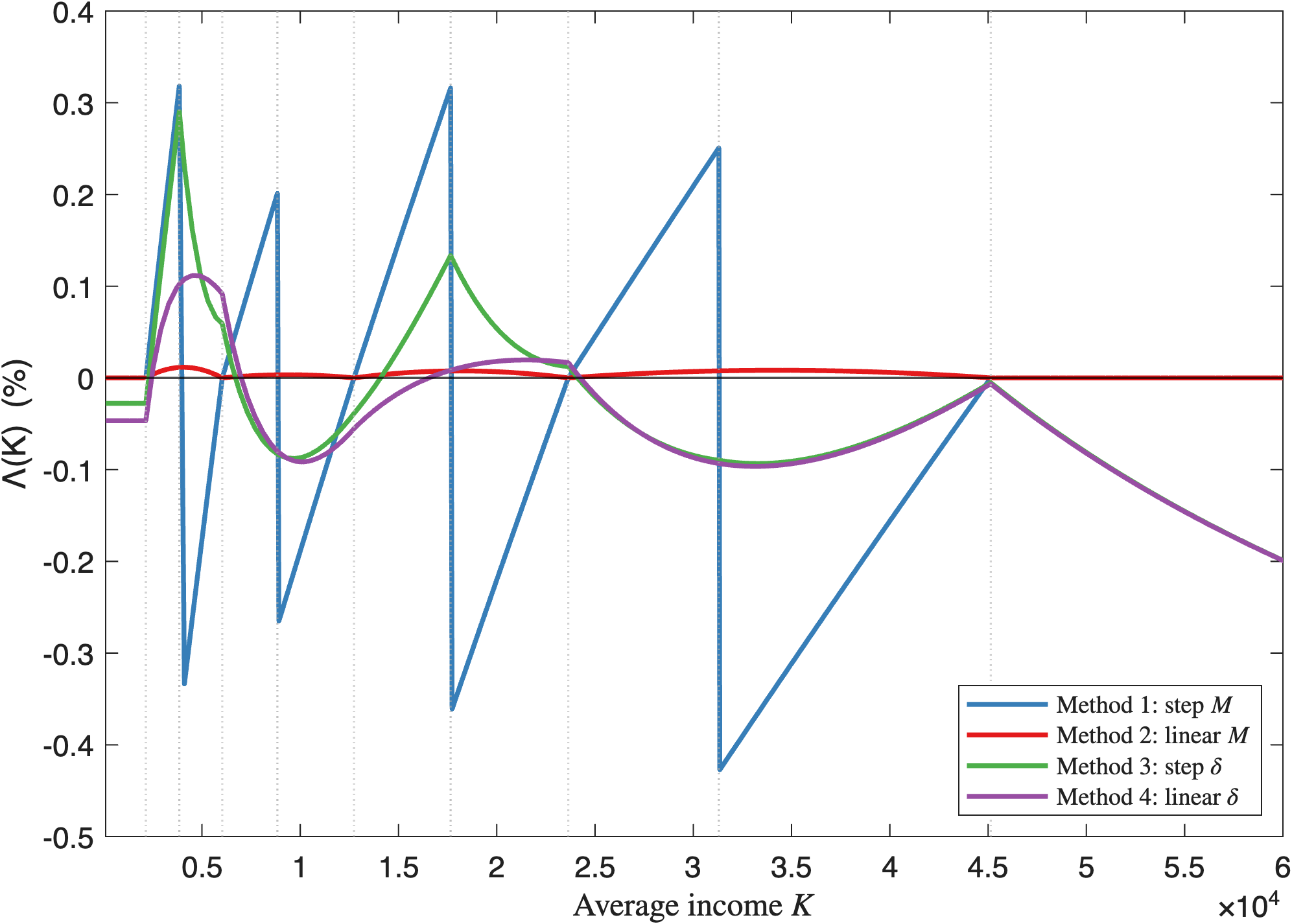}
    \caption{Residual subsidy rate after applying each of the four methods.}
    \label{fig:residual_subsidy}
\end{figure}

The differences across the four methods reflect the way each rule approximates the continuously fair benchmark. Method 1 produces the largest oscillations because its stepwise average divisor creates local over- and under-corrections within each income bracket. Method 2 has the smallest residual deviations because it directly interpolates the average counting month, which is the object entering the residual-subsidy formula. Methods 3 and 4 also keep the residual subsidy small, but their deviations are larger than Method 2 because these methods impose monotone marginal counting months and only approximate the fair anchor values. Method 3 generates a more segmented residual pattern because its marginal counting month is piecewise constant. Method 4 produces a smoother residual profile because its marginal counting month varies linearly across adjacent income anchors.

The comparison therefore illustrates a trade-off. Method 2 is the cleanest average-divisor approximation to the continuously fair benchmark. Methods 3 and 4 are marginal-rule implementations: they sacrifice exact average-divisor matching in order to obtain a monotone marginal conversion schedule. This feature is useful if policymakers prefer a rule that governs the conversion of each additional unit of career income, analogous to the way a marginal tax schedule governs the treatment of additional income. Importantly, all four methods achieve the main policy objective of sharply reducing the reverse transfer generated by the current official rule.

Finally, all quantitative results in this section are conditional on the valuation rate $r=7\%$.\footnote{The choice of 7\% roughly matches the average annual return of the National Social Security Fund from its establishment in 2000 to 2024, which is 7.39\%. Source: \url{https://english.news.cn/20250930/c9bf3340327e495e9b3beabb02deb24f/c.html}} The discount rate affects both the level of the annuity factor and the size of the mortality-based inequality. A higher discount rate lowers all annuity factors and reduces differences across income groups, because payments far in the future receive less weight in present value. More formally, if one income quintile has a remaining-lifetime distribution that first-order stochastically dominates that of another quintile (for example, when the non-crossover condition is imposed), then the annuity-factor gap between the two groups is decreasing in the discount rate; a short proof is given in Appendix~\ref{app:hermite_r}. Conversely, a lower discount rate raises the fair counting month and amplifies the reverse transfer embedded in an age-only divisor. The same logic applies when retirement benefits are indexed after retirement: positive benefit growth is equivalent to a lower effective discount rate and therefore magnifies the importance of differential longevity. The qualitative conclusion of this section is robust, but the quantitative magnitude of the subsidy should always be interpreted jointly with the assumed discount rate.

\section{Conclusion}
\label{sec:conclusion}

This paper studies how mortality heterogeneity across income groups affects actuarial fairness in China's notional defined contribution (NDC) pension system. We develop a mortality-differentiated Lee--Carter framework with a common period effect estimated from national mortality data and group-specific baseline schedules estimated from CHARLS. To model subgroup baseline mortality parsimoniously under limited data, we parameterize the baseline schedules using Hermite splines and impose simple shape restrictions.

Empirically, the current age-only annuity divisor is too low for all income groups under our valuation basis and yields larger subsidies to higher-income retirees. At retirement age 60 in 2020, the actuarially fair counting month ranges from 157.0 to 161.1 months, compared with the official value of 139. At age 63, the fair counting month ranges from 153.3 to 157.8 months, compared with the official value of 117. In both cases, the subsidy rises monotonically with income, implying not only an aggregate actuarial shortfall but also a reverse transfer from shorter-lived lower-income groups to longer-lived higher-income groups. Projections to 2040 suggest that this distortion is persistent and likely to widen as mortality continues to improve.

We then examine four implementable income-dependent annuitization rules. Even the simple bracket-based average-divisor rule substantially reduces the unfairness created by the current age-only rule, although it introduces threshold distortions. The piecewise-linear average-divisor rule provides the closest approximation to the continuously actuarially fair benchmark in our calibration. The two marginal-counting-month methods are useful when the policy objective is to govern the conversion of each additional unit of career income directly, much like a marginal tax schedule. These methods sacrifice exact average-divisor matching in order to obtain monotone marginal conversion schedules, but their residual distortions remain small relative to the unreformed official rule.

Two broader conclusions follow. First, actuarial fairness in NDC pension design cannot be evaluated using aggregate mortality alone; mortality heterogeneity matters both for distributional incidence and for the level of the annuity divisor. Second, parsimonious multi-source mortality models can make these questions empirically tractable even when subgroup data are sparse. Future work could extend the framework by incorporating richer subgroup data, allowing group-specific period effects when longer panels become available, and embedding the annuitization reform in a fuller model of notional accumulation, retirement behavior, and policy incidence.

\section*{Data availability statement}
The national mortality data of China are sourced from the official yearbooks published by the National Bureau of Statistics. The China Health and Retirement Longitudinal Study (CHARLS) data are publicly available for registered users at \url{https://charls.pku.edu.cn/en/}. The analysis code is available from the authors upon reasonable request.

\section*{Funding statement}

Kenneth Q. Zhou acknowledges the support of the Natural Sciences and Engineering Research Council of Canada (NSERC), [RGPIN-2025-04157] and [DGECR-2025-00488]. Xiaobai Zhu acknowledges the support of the National Natural Science Foundation of China [No. 12301613] and the Shanghai Sci-tech Co-research Program [No. 25HB2702200].

\section*{Conflict of interest disclosure}

The authors declare no conflicts of interest.

\section*{AI usage disclosure}
Portions of this manuscript were drafted or edited with AI-assisted writing tools. All content was reviewed, verified, and approved by the authors.

\bibliography{reference}
\bibliographystyle{apalike}

\appendix

\section{Official Counting Months} \label{sec:month}
\begin{table}[H]
    \centering
    \begin{tabular}{cc|cc|cc}
    \toprule
       Age ($x$) & $M_x^{\mathrm{off}}$ & Age ($x$) & $M_x^{\mathrm{off}}$ & Age ($x$) & $M_x^{\mathrm{off}}$ \\
    \midrule
        40 & 233 & 50 & 195 & 60 & 139\\
        41 & 230 & 51 & 190 & 61 & 132\\
        42 & 226 & 52 & 185 & 62 & 125\\
        43 & 223 & 53 & 180 & 63 & 117\\
        44 & 220 & 54 & 175 & 64 & 109\\
        45 & 216 & 55 & 170 & 65 & 101\\
        46 & 212 & 56 & 164 & 66 & 93\\
        47 & 207 & 57 & 158 & 67 & 84\\
        48 & 204 & 58 & 152 & 68 & 75\\ 
        49 & 199 & 59 & 145 & 69 & 65\\
           &     &    &     & 70 & 56\\
    \bottomrule 
    \end{tabular}
    \caption{Current official counting months in China's notional defined contribution (NDC) pension system.}
    \label{tab:China_annuity_divisor}
\end{table}

\section{Discount Rates and Annuity-Factor Gaps Across Income Groups}\label{app:hermite_r}

This appendix establishes the claim used in Section~4 that, if one income group has a remaining-lifetime distribution that first-order stochastically dominates that of another group, then the annuity-factor gap between the two groups decreases with the discount rate.

Let \({}_tp_{x,j}=P(T_{x,j}>t)\) denote the survival probability from age \(x\) to age \(x+t\) for income group \(j\), where \(T_{x,j}\) is remaining lifetime. Suppose that group \(j\) is longer-lived than group \(i\) in the sense that
\[
{}_tp_{x,j}\ge {}_tp_{x,i},
\qquad
\forall t\ge 0.
\]

The monthly annuity-due factor for group \(j\) can be written as
\[
\ddot{a}^{(12)}_{x,j}(r)
=
\frac{1}{12}\sum_{m=0}^{\infty} v^{m/12}\,{}_{m/12}p_{x,j},
\qquad
v=(1+r)^{-1}.
\]
Hence the gap between groups \(j\) and \(i\) is
\[
\Delta_{ij}(r)
=
\ddot{a}^{(12)}_{x,j}(r)-\ddot{a}^{(12)}_{x,i}(r)
=
\frac{1}{12}\sum_{m=0}^{\infty} v^{m/12}
\left(
{}_{m/12}p_{x,j}-{}_{m/12}p_{x,i}
\right).
\]
By assumption, each survival-probability difference in parentheses is nonnegative.

Differentiating term by term with respect to \(r\), we obtain
\[
\Delta_{ij}'(r)
=
\frac{1}{12}\sum_{m=1}^{\infty}
\frac{\partial v^{m/12}}{\partial r}
\left(
{}_{m/12}p_{x,j}-{}_{m/12}p_{x,i}
\right).
\]
Since
\[
\frac{\partial v^{m/12}}{\partial r}
=
-\frac{m}{12}(1+r)^{-m/12-1}
<0
\qquad
\text{for all } m\ge 1,
\]
and the survival-probability differences are nonnegative, every term in the sum is nonpositive. Therefore,
\[
\Delta_{ij}'(r)\le 0.
\]

Thus, the annuity-factor gap between the longer-lived and shorter-lived group decreases as the discount rate increases. Under the bounded survival probabilities used here, term-by-term differentiation is justified by dominated convergence. \qed

\end{document}